\documentclass[useAMS,usenatbib]{mn2e}
\usepackage{graphicx}
\def\aj{AJ}%
\def\araa{ARA\&A}%
\def\apj{ApJ}%
\def\apjl{ApJ}%
\def\apjs{ApJS}%
\def\aap{A\&A}%
\def\mnras{MNRAS}%
\def\pasp{PASP}%
\def\nat{Nature}%
\title[Low-mass star formation by disc fragmentation]
{The properties of brown dwarfs and low-mass hydrogen-burning stars formed by disc fragmentation
}
\author[D. Stamatellos \& A.~P. Whitworth]
{Dimitris Stamatellos\thanks{E-mail:D.Stamatellos@astro.cf.ac.uk} and Anthony~P. Whitworth \\ 
School of Physics \& Astronomy,Cardiff University, Cardiff, CF24 3AA, Wales, UK}

\begin{document}

\date{Accepted 2008 August 15. Received 2008 July 24; in original form 2008 July 24}

\pagerange{\pageref{firstpage}--\pageref{lastpage}} \pubyear{2007}

\maketitle

\label{firstpage}

\begin{abstract}
We suggest that a high proportion of brown dwarf stars are formed by gravitational fragmentation of massive extended discs around Sun-like primary stars. We argue that such discs should arise frequently, but should be observed infrequently, precisely because they fragment rapidly. By performing an ensemble of radiation-hydrodynamic simulations, we show that such discs typically fragment within a few thousand years, and produce mainly brown dwarf (BD) stars, but also planetary-mass (PM) stars and very low-mass hydrogen-burning (HB) stars. Subsequently most of the lower-mass stars (i.e. the PM and BD stars) are ejected by mutual interactions. We analyse the statistical properties of these stars, and compare them with observations.

After a few hundred thousand years the Sun-like primary is typically left with a close low-mass HB companion, and two much wider companions: a low-mass HB star and a BD star, or a BD-BD binary. The orbits of these companions are highly eccentric, and not necessarily coplanar, either with one another, or with the original disc. There is a BD desert extending out to at least $\,\sim\!100\,{\rm AU}$; this is because BDs tend to be formed further out than low-mass HB stars, and then they tend to be scattered even further out, or even into the field.

BDs form with discs of a few Jupiter masses and radii of a few tens of AU, and they are more likely to retain these discs if they remain bound to the primary star.  Binaries form by pairing of the newly-formed stars in the disc, giving a low-mass binary fraction of $\,\sim\!0.16$. These binaries include close and wide BD/BD binaries and BD/PM binaries. Binaries can be ejected into the field and survive, even if they have quite wide separations. BDs that remain as companions to Sun-like stars are more likely to be in BD/BD binaries than are BDs ejected into the field. The presence of close and distant companions around Sun-like stars may inhibit planet formation by core accretion.

We conclude that disc fragmentation is a robust mechanism for BD formation. Even if only a small fraction of Sun-like stars host the required massive extended discs, this mechanism can produce all the PM stars observed, most of the BD stars, and a significant proportion of the very low-mass HB stars.

\end{abstract}


\section{Introduction}
\label{SEC:INTRO}

Brown dwarfs were predicted theoretically by Kumar (1962) and Hayashi \& Nakano (1963). They were first detected more than thirty years later (Rebolo et al. 1995; Nakajima et al. 1995; Oppenheimer et al. 1995). Since then hundreds have been observed, both in nearby young star clusters and in the field (see Luhman et al. 2007 and references therein).

%%%%%%%%%%%%%%
\begin{table}
\begin{minipage}{0.45\textwidth}
\caption{Nomenclature.}
\label{TAB:NOMEN}
\centering
\renewcommand{\footnoterule}{}  % to avoid a line before footnotes
\begin{tabular}{@{}ll} \hline
\noalign{\smallskip}
{\sc term} & {\sc meaning} \\
\noalign{\smallskip}
\hline
\noalign{\smallskip}
star & pressure-supported object forming on a \\
 & \hspace{0.2cm} dynamical timescale, by gravitational instability. \\
planet & pressure-supported object forming on a much \\
 & \hspace{0.2cm} longer than dynamical timescale, by core \\
 & \hspace{0.2cm} accretion. \\\hline
HB & hydrogen-burning star, \hspace{1.70cm} $M_\star\geq 80\,{\rm M}_{_{\rm J}}$. \\
BD & brown-dwarf star, \hspace{1.20cm} $13\,{\rm M}_{_{\rm J}}\leq M_\star\leq 80\,{\rm M}_{_{\rm J}}$. \\
PM & planetary-mass star, \hspace{2.05cm} $M_\star\leq 13\,{\rm M}_{_{\rm J}}$. \\\hline
ur-star & star at the centre of the ur-disc; present from \\
 & \hspace{0.2cm} the start of the simulation, and provides the \\
 & \hspace{0.2cm} background gravitational and radiation fields. \\
ur-disc & disc around the ur-star; the material whose \\
 & \hspace{0.2cm} dynamics is to be simulated. \\
ur-system & ur-star plus ur-disc. \\
\noalign{\smallskip}
\hline
\end{tabular}
\end{minipage}
\end{table}
%%%%%%%%%%%

A distinction is often made between brown dwarfs below, and stars above, the hydrogen-burning limit at $\sim 80\,{\rm M}_{_{\rm J}}$ (where ${\rm M}_{_{\rm J}}=2\times 10^{30}\,{\rm g}$ is the mass of Jupiter). Similarly, a distinction is often made between planetary-mass objects below, and brown dwarfs above, the deuterium-burning limit at $\sim 13\,{\rm M}_{_{\rm J}}$. However, in the context of star formation, these distinctions are -- at best -- unhelpful.

We will therefore refer collectively to all objects forming by gravitational instability, on a dynamical timescale, as {\it stars}. Within this definition, we will label as hydrogen-burning stars (HBs) those stars with masses above the H-burning limit, as brown dwarf stars (BDs) those stars with masses between the H-burning limit and the D-burning limit, and as planetary-mass stars (PMs) those stars with masses below the D-burning limit. This nomenclature is summarised in Table \ref{TAB:NOMEN}. Since there are only 3 PMs formed in our simulations, we will, for statistical purposes, sometimes lump the PMs in with the BDs.

We note that for very low mass stars the dynamical timescale is essentially the freefall timescale at rather high density. Therefore it is very short, specifically 
\begin{eqnarray}
t_{_{\rm FF}}&\simeq&\frac{G\,M}{10\,c_{_{\rm S}}^3}\;\simeq\;0.06\,{\rm kyr}\,\left(\frac{M}{{\rm M}_{_{\rm J}}}\right)\,,
\end{eqnarray}
where we have estimated the freefall time at the centre of a marginally unstable isothermal sphere, and $c_{_{\rm S}}=0.2\,{\rm km}\,{\rm s}^{-1}$ is the isothermal sound speed for molecular gas at $T\sim 10\,{\rm K}$. Therefore the formation of PMs, BDs and low-mass HBs (say below $\sim 200\,{\rm M}_{_{\rm J}}$) is extremely rapid ($t_{_{\rm FF}}\la 10\,{\rm kyr}$).

In contrast to our definition of stars, we will refer to all objects which form by core accretion, on a much longer timescale ($\ga 1000\,{\rm kyr}$), as planets.

With these definitions, we can discuss a single coherent Stellar Initial Mass Function (IMF) which extends below the H- and D-burning limits. This stellar IMF is bounded by a minimum stellar mass, which is determined by the requirement that a star-forming condensation is able to radiate away, on a dynamical timescale, the $PdV$ work done by compression. This requirement is usually called the opacity limit, but see Masunaga \& Inutsuka (1999). For contemporary star formation, in the solar vicinity, the minimum stellar mass is estimated to lie in the range from $1\;{\rm to}\;7\,{\rm M}_{_{\rm J}}$  (Rees 1976; Low \& Lynden-Bell 1976; Boyd \& Whitworth 2005; Whitworth \& Stamatellos 2006).

Given this low minimum stellar mass, it is very likely that the Stellar IMF overlaps the Planetary Mass Function in the interval between 1 and $10\,{\rm M}_{_{\rm J}}$. However, with the definitions above, the two mass functions are distinct in their origin.

During the last decade there has been much speculation about the origin of BDs (see review by Whitworth et al. 2007). Currently there are three main theories proposed: (i) magneto-turbulent or gravo-turbulent fragmentation, (ii) premature ejection of protostellar embryos from their natal cores, and (iii) disc fragmentation. 

Padoan \& Nordlund (2004) suggest that magneto-turbulent fragmentation of molecular clouds produces prestellar cores of brown-dwarf mass with densities high enough for them to be gravitationally unstable and collapse. Their model follows the standard paradigm of low-mass star formation, i.e. a prestellar core collapses to form a single star. The model does not attempt to explain the formation of BDs in clusters and binaries. Additionally, the highly co-ordinated supersonic velocities needed for these dense low-mass prestellar cores to form have yet to be observed.

Gravo-turbulent fragmentation of collapsing prestellar cores may also produce BDs. Simulations (e.g. Bate et al. 2003; Goodwin et al. 2004a,b, 2006) of collapsing turbulent cores produce both HBs and BDs, with numbers in good agreement with observation. However, these simulations have been performed with a barotropic equation of state, and therefore the effects of radiative transfer (e.g. Boss et al. 2000; Whitehouse \& Bate 2006; Stamatellos et al. 2007a) are not properly captured.

The ejection scenario is closely associated with the gravo-turbulent fragmentation mechanism. In numerical simulations of collapsing turbulent cores, BDs frequently form when protostellar embryos (i.e. very low-mass protostars which have just formed) are ejected from their natal cores due to dynamical interactions with other protostars. As a consequence of ejection they stop accreting, and so they never acquire sufficient mass to sustain hydrogen burning (Reipurth \& Clarke 2001). 

BDs can also form by disc fragmentation (e.g. Whitworth \& Stamatellos 2006; Stamatellos et al. 2007b). There are two conditions which must be met for a disc to fragment. First, the disc must be massive enough so that locally gravity can overcome thermal and centrifugal support (Toomre 1964), i.e. 
\begin{equation}
Q(R)\equiv\frac{c(R)\,\kappa(R)}{\pi\,G\,\Sigma(R)}\;\la\;1\,;
\end{equation}
here $Q$ is the Toomre parameter, $c$ is the sound speed, $\kappa$ is the epicyclic frequency, and $\Sigma$ is the surface density. Second, the disc must cool efficiently so that proto-condensations forming in the disc do not simply undergo an adiabatic bounce and dissolve. Both theory and simulations (Gammie 2001; Johnson \& Gammie 2003; Rice et al. 2003, 2005;  Mayer et al. 2004; Mejia et al. 2005; Stamatellos et al. 2007b) indicate that for disc fragmentation to occur, the cooling time must satisfy
\begin{equation}
t_{_{\rm COOL}}\;\,<\;\;{\cal C}(\gamma)\,t_{_{\rm ORB}}\,,\hspace{1.0cm}0.5\la{\cal C}(\gamma)\la 2.0\,;
\end{equation}
here $t_{_{\rm ORB}}$ is the local orbital period, and $\gamma$  the adiabatic exponent. Numerical simulations suggest that disc fragmentation produces BDs, PMs and also low-mass HBs (Stamatellos et al. 2007b). These objects may either remain bound to the primary star, or be ejected into the field. The main concern with this model is whether massive extended discs actually form in the first place. Observations of such discs are difficult, since they are short lived and heavily embedded in their parental clouds. We return to this issue below.

Two other mechanisms that might form BDs are the photo-erosion of massive prestellar cores by nearby ionising stars (Whitworth \& Zinnecker 2004), and fragmentation during the second collapse, i.e. when H$_2$ dissociates (Whitworth \& Stamatellos 2006). The former mechanism can only operate in the vicinity of a massive star, and therefore it cannot be a major source of BDs. The latter mechanism has been advocated on theoretical grounds, but it has not yet been shown to work by numerical simulations (e.g. Bate 1998; Stamatellos et al. 2007a).

The above BD formation mechanisms are not exclusive of each other and can work in conjunction in star forming regions. For example, gravo-turbulent fragmentation can generate cores which collapse to form discs, the discs may then fragment to produce BDs, and the BDs may avoid accreting additional mass by being ejected.

Since all these mechanisms should produce both BDs and low-mass HBs, we expect there to be a continuity in their statistical properties across the H-burning limit, and this is indeed what seems to be observed. Like low-mass HBs, young BDs have discs (and all the consequences thereof, i.e. IR excesses, signatures of magnetospheric accretion, X-rays, outflows, etc.), they are sometimes found in binary and higher multiple systems, and they coexist with stars (e.g. Luhman et al. 2007 and references therein).

Nonetheless, some authors (e.g. Thies \& Kroupa 2007) find observational evidence that BDs form in a different way to HBs. In this regard the most critical factor is the {\it brown dwarf desert}, i.e. the lack of close BD companions to Sun-like stars. Radial velocity surveys have revealed a large number of planets orbiting close to Sun-like stars, as well as a large number of low-mass HBs, but very few BDs (Marcy \& Butler 2000; see Burgasser et a. 2007 for review). As we shall show below, the brown dwarf desert finds a natural explanation in the mechanics of disc fragmentation and subsequent protostellar dynamics.

In this paper we explore the fragmentation of a disc having comparable mass to the central primary star. In Section 2 we discuss the numerical methods we use, and in Section 3 we describe the initial disc configuration. In Section 4 we present in detail one of the simulations performed, and then describe an ensemble of simulations. In Section 5 we discuss the statistical properties of the stars produced by disc fragmentation and compare them with the observed properties of BDs and low-mass HBs. In Section 6 we describe what observers might expect to see if they observe a fragmented disc after a few hundred years, and in Section 7 we discuss the implications of the disc fragmentation mechanism for planet formation. Finally, in Section 8 we summarise our results. 

\section{Numerical method}

The evolution of the disc is initially followed using Smoothed Particle Hydrodynamics (SPH), until $70\;{\rm to}\;80\%$ of the disc mass has been accreted, either onto the stars condensing out of the disc, or onto the central primary star; this typically takes 10 to 20  kyr. Then the residual gas is ignored and the long term dynamical evolution of the resulting ensemble of stars is followed up to 200 kyr, using an N-body code.

For the hydrodynamics we use the SPH code {\sc dragon} (e.g. Goodwin et al. 2004a,b), which invokes an octal tree (to compute gravity and find neighbours), adaptive smoothing lengths, multiple particle timesteps, and a second-order Runge-Kutta integration scheme. The code uses time-dependent viscosity with parameters $\alpha^\star=0.1$, $\beta=2\alpha$  (Morris \& Monaghan 1997) and a Balsara switch (Balsara 1995), so as to reduce artificial shear viscosity (Artymowicz \& Lubow 1994; Lodato \& Rice 2004; Rice, Lodato, \& Armitage 2005). The smoothing lengths are adapted so that each particle has exactly ${\cal N}_{_{\rm NEIB}}=50$ neighbours; this reduces numerical diffusion to a minimum (Attwood, Goodwin \& Whitworth 2007). Sinks are created wherever a bound condensation forms and its density exceeds $\rho_{_{\rm SINK}}=10^{-9}\,{\rm g}\,{\rm cm}^{-3}$. SPH particles are accreted onto a sink if they are within $R_{_{\rm SINK}}=1\,{\rm AU}$ of the sink and bound to it. Sinks are identified with stars. We note that our high value of $\rho_{_{\rm SINK}}$ makes the use of sinks relatively safe.

The energy equation and associated radiative transfer are treated with the method of Stamatellos et al. (2007a). This method takes into account compressive heating or expansive cooling, viscous heating, radiative heating by the background, and radiative cooling.   The method does not solve the radiation transfer equation (which is unfeasible in hydrodynamic simulations with the current computing resources), but it is an appoximate method that performs well, in the optically-thin, intermediate and optically-thick regimes, and has been extensively tested (Stamatellos et al. 2007a). In particular it reproduces the detailed 3D results of Masunaga \& Inutsuka (2000), Boss \& Bodenheimer (1979), Boss \& Myhill (1992), Whitehouse \& Bate (2006),  and also the analytic test of Spiegel (1957). It also performs well in the analytic test of Hubeny (1990) for disc geometries. It is easy to implement in particle- and grid-based codes, and highly efficient (incurring only a $3\%$ overhead, in comparison with simulations performed using a barotropic equation of state). 

The gas is assumed to be a mixture of hydrogen and helium. We use an equation of state (Black \& Bodenheimer 1975; Masunaga et al. 1998; Boley et al. 2007) that accounts for the rotational and vibrational degrees of freedom of molecular hydrogen, and for the different chemical states of hydrogen and helium. We assume that ortho- and para-hydrogen are in equilibrium.

For the dust and gas opacity we use the parameterization  by Bell \& Lin (1994), $\kappa(\rho,T)=\kappa_0\ \rho^a\ T^b\,$, where $\kappa_0$, $a$, $b$ are constants that depend on the species and the physical processes contributing to the opacity at each $\rho$ and $T$. The opacity changes due to ice mantle evaporation and the sublimation of dust are taken into account, along with the opacity contributions from molecules and H$^-$ ions. 

For the N-body part of the simulation, exploring the long term dynamical evolution of the stars formed by fragmentation of the outer disc (from $\sim 20 \;{\rm to}\;200 \,{\rm kyr}$), we use a 4th-order Hermite integration scheme (Makino \& Aarseth 1992), with a conservative timestep criterion so that energy is conserved to better than one part in $10^8$ (Hubber \& Whitworth 2005).

\section{Disc Initial conditions}

Only a few extended massive discs have been observed around Sun-like stars (e.g. Eisner et al. 2005, 2008; Eisner \& Carpenter 2006; Rodriguez et al. 2005). However, we suggest that this is because the outer parts of such discs are rapidly dissipated by gravitational fragmentation -- rather than because they seldom form in the first place. Indeed, it seems that the formation of such discs is inevitable.

For example, a $1.4\,{\rm M}_\odot$ prestellar core with ratio of rotational to gravitational energy $\beta\equiv{\cal R}/|\Omega|$ will -- if it collapses monolithically -- forms a protostellar disc with outer radius $R_{_{\rm DISC}}\sim 400\,{\rm AU}\,(\beta/0.01)$. Since the observations of Goodman et al. (1993) indicate that many prestellar cores have $\beta \sim 0.02$, the formation of extended discs should be rather common.

Alternatively, if an existing $0.7\,{\rm M}_\odot$ protostar attempts to assimilate matter with specific angular momentum $h$, this matter is initially parked in an orbit at $R_{_{\rm ORBIT}}\sim 400\,{\rm AU}\,(h/5\times 10^{20}\,{\rm cm}^2\,{\rm s}^{-1})^2$. The quantum of specific angular momentum used to normalise this relation is rather modest by protostellar standards, as can be seen by expressing it in terms of a lever-arm in parsecs and a velocity in kilometres per second, viz. $5\times 10^{20}\,{\rm cm}^2\,{\rm s}^{-1}\equiv (0.02\,{\rm pc})\,\times\,(0.1\,{\rm km}\,{\rm s}^{-1})$.

Previous simulations (Stamatellos et al.  2007b) suggest that the outer parts of massive extended discs fragment on a dynamical timescale ($\la 3 \,{\rm kyr}$), so they are indeed very short-lived. 

We shall assume a star-disc system (hereafter the {\it ur-system}), in which the central primary star (hereafter the {\it ur-star}) has initial mass $M_1=0.7\,{\rm M}_{\sun}$. Initially the disc (hereafter the {\it ur-disc}) has mass $M_{_{\rm D}}=0.7\,{\rm M}_{\sun}$, inner radius $R_{_{\rm IN}}=40\,{\rm AU}$, outer radius $R_{_{\rm OUT}}=400\,{\rm AU}$, surface density 
\begin{equation}
\Sigma_{_0}(R)=\frac{0.03\,{\rm M}_{\sun}}{{\rm AU}^2}\,\left(\frac{R}{\rm AU}\right)^{-7/4}\,,
\end{equation}
temperature
\begin{equation}\label{EQN:TBG}
T_{_0}(R)=250\,{\rm K}\,\left(\frac{R}{\rm AU}\right)^{-1/2}+10~{\rm  K}\,,
\end{equation}
and hence approximately uniform initial Toomre parameter $Q\sim 0.9$. Thus the disc is at the outset marginally gravitationally unstable. (The {\it ur-} prefix is used solely because it otherwise becomes difficult later on to discuss unambiguously the smaller discs that attend newly-formed BDs, the orbital parameters of BD/BD binaries, and so forth).  We have designed a disc with $Q<1$ over a wide range of radii, because we are wanting to establish which parts of the disc can, and which cannot, fragment because they obey, or violate, the Gammie criterion.

The radiation of the ur-star is taken into account by invoking a background blackbody radiation field with temperature $T_{_0}(R)$. In effect, this means that, if the material in the disc is heated by compression and/or viscous dissipation, it can only cool radiatively if it is warmer than $T_{_0}(R)$ given by Eqn. (\ref{EQN:TBG}).

\section{Simulations: Overview}

We perform twelve simulations: ten using 150,000 particles, one with 250,000 and one with 400,000 (the last two to check convergence). The minimum resolvable mass (corresponding to 100 SPH particles) is $\la 0.5\,{\rm M}_{_{\rm J}}$, and therefore the Jeans condition is obeyed at all times. 

All of the simulations start with the same ur-system (i.e. the same ur-disc and ur-star parameters). The only difference (apart from the number of particles in the last two simulations) is the random seed used to construct each ur-disc. Thus the Poisson density fluctuations are different in each simulation. This allows us to study the statistical properties of the stars produced. 

%%%%%%%%%%%%%%%
\begin{figure*}
\centerline{
\includegraphics[height=17cm,angle=-90]{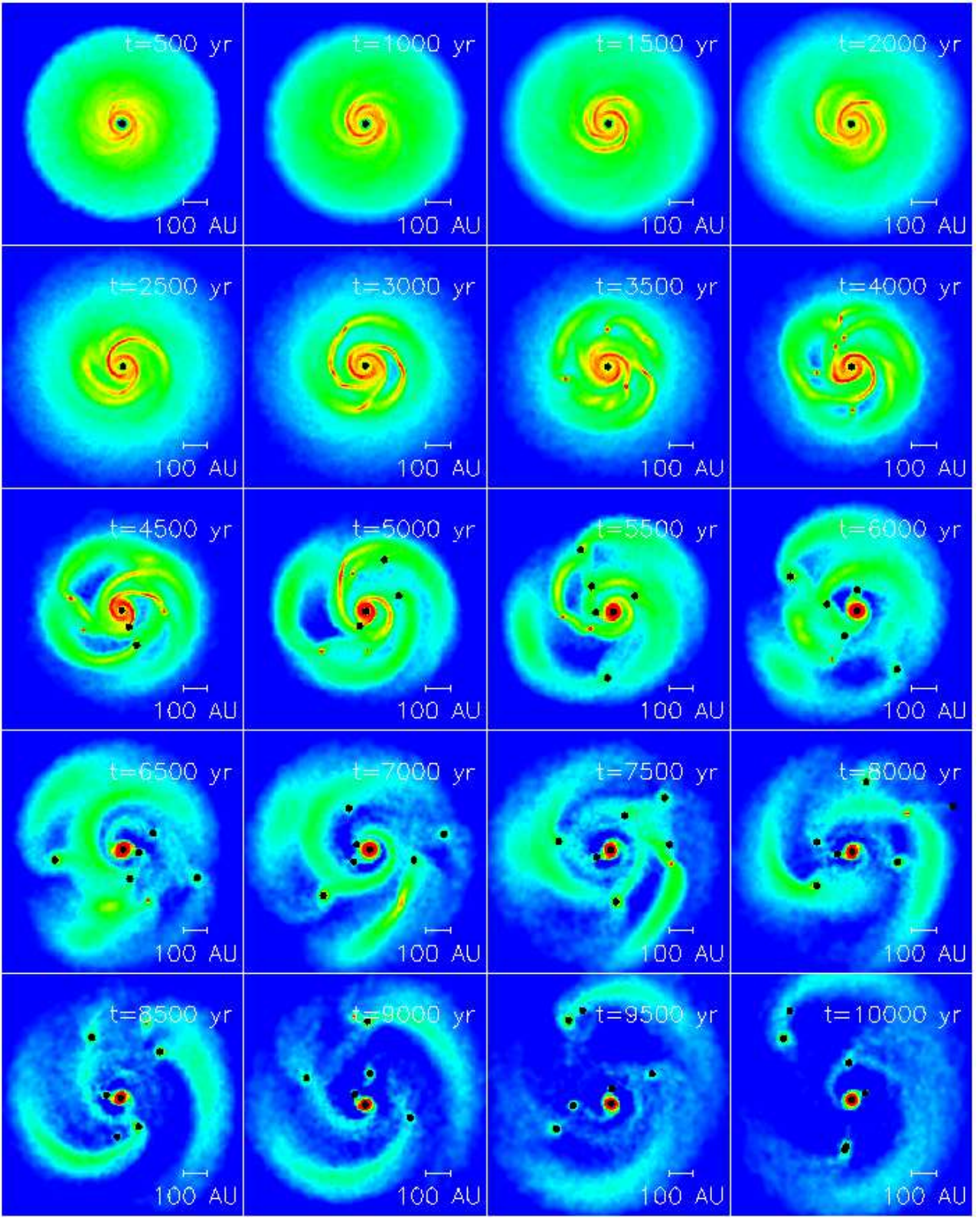}
}
\caption{Radiative hydrodynamic simulation of the evolution of  a 0.7-$M_{\sun}$ ur-disc around a 0.7-$M_{\sun}$ ur-star. Snapshots of the logarithm of the column density are presented from 0.5 to 10~kyr, every 0.5 kyr (as marked on each graph). Such ur-discs are gravitationally unstable and they cool efficiently. Hence they quickly fragment to form BDs and low-manss HBs. This particular simulation produces 4 BDs and 3 low-mass HBs. After 200 kyr, 4 of these stars have been ejected into the field.}
\label{fig:rdm1}
\end{figure*}
%%%%%%%%%%%%%

Because the ur-discs are Toomre unstable, spiral arms form and sweep up material into dense condensations. Within a few thousand years, between 5 and 11 stars have condensed out of the disc. In all 12 simulations a total of 96 stars are formed. 3\% are PMs, 67\% are BDs, and the rest are low-mass HBs (with masses up to $\sim 200\,{\rm M}_{_{\rm J}}\equiv 0.2\,{\rm M}_\odot$). (Hereforth BD statistics will include PMs, unless otherwise  stated).

Stars initially condense out of the ur-disc where $Q<1$ and $t_{_{\rm COOL}}<0.5\ t_{_{\rm ORB}}$, i.e. typically at distances $R\ga 100\,{\rm AU}$ AU from the central star. HBs generally form closest to the ur-star, and PMs form furthest from the ur-star. There are three factors producing this trend of lower-mass stars forming further out in the ur-disc: (i) stars condensing out at small $R$ condense out earlier, so they then have longer to accrete; (ii) the ambient density at small $R$ is higher, so there is more matter to accrete; and (iii) the most massive stars tend to migrate inwards, so they get to mop up the inner parts of the ur-disc where no stars have condensed out. 

With between 5 and 11 stars formed in each ur-disc, initially between $\sim 60$ and $\sim 400\,{\rm AU}$, interactions between the stars are frequent. During 2-body interactions, binary systems can be formed if there is sufficient dissipation (e.g. due to the small accretion discs which normally attend the individual stars). In 3-body interactions, the lowest-mass star is often ejected into the field, and therefore stops accreting altogether. $55\%$ of the stars formed end up being ejected into the field. Most of the stars below the H-burning limit, and all the stars below the D-burning limit are ejected into the field.

Most are ejected as single stars, but some are ejected as components of binary systems. These binary systems comprise BD/BD pairings, BD/HB pairings, and even one BD/PM pairing; there are both wide and close systems.

Ejected BDs frequently retain accretion discs with masses on the order of a few Jupiter masses. Hence they can sustain accretion and outflows as observed (Jayawardhana et al. 2003; Muzerolle et al. 2005; Mohanty et al. 2005; Whelan et al. 2005). 

The final outcome typically has a low-mass HB orbiting the ur-star at close distance ($R\la 10\;{\rm to}\;20\,{\rm AU}$), a low-mass HB in a wider orbit, and a couple of BDs or a BD/BD binary orbiting at large distance ($R\ga 100\;{\rm to}\;200\,{\rm AU}$). All the other stars have been ejected into the field. 

In Fig.~\ref{fig:rdm1} we present column-density images of the ur-disc, every 500 yr, for one of the simulations. In this simulation 7 stars form: 4 BDs and 3 low-mass HBs. At the end of the simulation (i.e. after the hydrodynamic simulation {\it and} the N-body simulation, at $200\,{\rm kyr}$), only 3 stars remain bound to the ur-star: a low-mass HB orbiting close to the ur-star ($R\sim 10\,{\rm AU}$), and a binary comprising a BD and a low-mass HB, orbiting at $R\sim 7,700\,{\rm AU}$. 

\begin{figure}
\centerline{
\includegraphics[height=11cm]{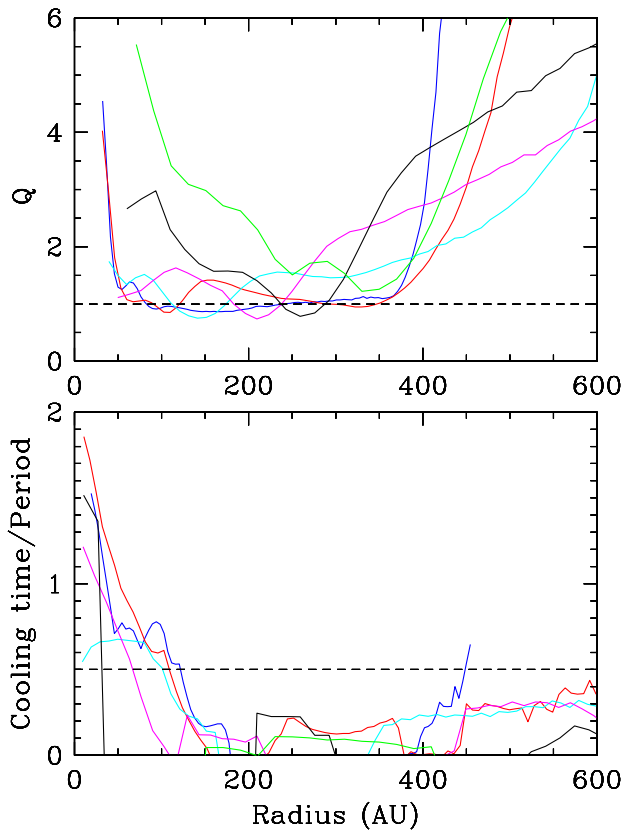}}
\caption{The Toomre parameter $Q$ and the ratio of the cooling time to the orbital period $t_{_{\rm COOL}}/t_{_{\rm ORB}}$ (both azimuthaly averaged) are plotted against distance from the ur-star, every 1 kyr from 0 to $6 \,{\rm kyr}$ (blue, red, cyan, magenta, black, green).}
\label{fig:rdm1.qtoomcool}
\end{figure}

Fig.~\ref{fig:rdm1.qtoomcool} shows the Toomre parameter $Q$ and the ratio of the cooling time to the orbital time $t_{_{\rm COOL}}/t_{_{\rm ORB}}$ (both azimuthally averaged), every 1 kyr from 0 to 6 kyr. Initially the ur-disc is marginally unstable (blue line), and outside $100\,{\rm AU}$ the cooling time is favorable for fragmentation (i.e. $t_{_{\rm COOL}}<0.5\ t_{\rm ORB}$). Hence the ur-disc fragments here. As time progresses the region where $Q<1$ moves farther away from the ur-star as the gas in the inner regions is used up.

%%%%%%%%%%%%%%
%%%%%%%%%%%%

In the two simulations with higher resolution, the growth of gravitational instabilities, and the properties of the stars formed as a result of these instabilities, follow the same patterns as in the simulations with lower resolution. The variance is the same as between two simulations performed with the same resolution, and is attributable to the different seed noise in the different simulations (due to random positioning of the particles) and the chaotic, non-linear nature of gravitational instability. Thus, the simulations appear to be converged, in a statistical sense.

In the following sections we describe in detail the results of the simulations, focusing on the statistical properties of the stars formed, and compare them with the observed properties of BDs and low-mass HBs.

\section{Statistical properties of the low-mass stars produced by disc fragmentation}

\subsection {Formation time}

\begin{figure}
\centerline{
\includegraphics[height=0.5\textwidth]{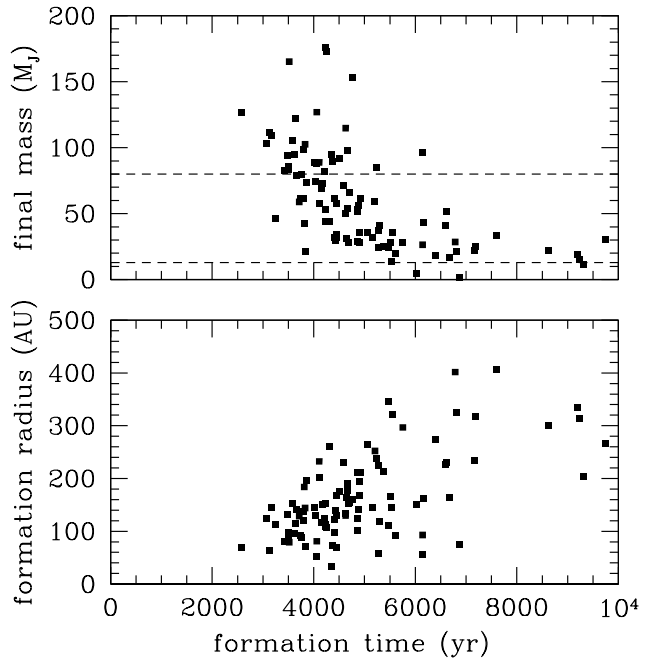}}
\caption{The final mass and formation radius of the stars formed by disc fragmentation plotted against formation time (top and bottom respectively).  Stars form between 2 and 10 kyr. In general, more massive stars form earlier and closer to the ur-star, whereas lower-mass stars form later and farther from the ur-star. }
\label{fig:tformation}
\end{figure}

The ur-discs start fragmenting after $2\,{\rm kyr}$, which corresponds to about one quarter of an outer rotation period ($1~{\rm ORP}=8\,{\rm kyr}$) (see Fig.~\ref{fig:tformation}). Stars condense out on a dynamical timescale. Hence star formation occurs faster in the inner ur-disc, but even at the edge of the ur-disc it is virtually over by $1\,{\rm ORP}$. The stars which form first, in the inner ur-disc, have more time to accrete, and more material to accrete, and hence they tend to end up as HBs, with masses up to $\sim 200\,{\rm M}_{\rm J}$ (see Fig.~\ref{fig:tformation}). Lower mass stars (i.e. BDs and PMs) tend to form at later times and at larger distances from the ur-star.

%%%%%%%%%%%%%%
%%%%%%%%%%%%

\subsection {Mass distribution}

\begin{figure}
\centerline{
\includegraphics[height=0.5\textwidth,angle=-90]{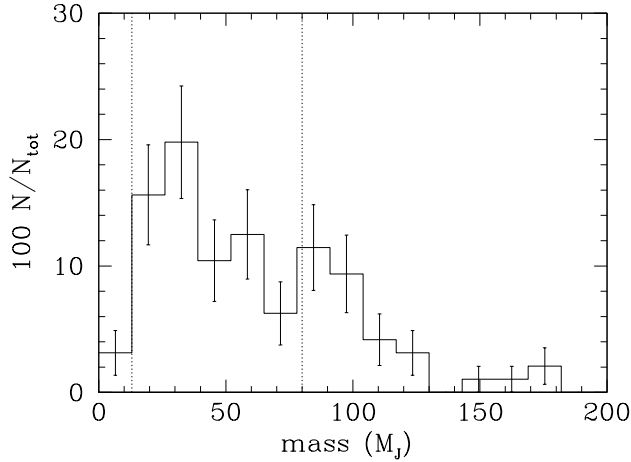}}
\caption{Mass spectrum of the stars produced by disc fragmentation. Most of the stars are BDs ($70\%$); the rest are low-mass HBs. The perpendicular dotted lines correspond to the D-burning limit ($\sim 13 M_{\rm J}$) and the H-burning limit ($\sim 80 M_{\rm J}$). The mass used for the histogram is the mass that the star has at the end of the SPH simulation. Most of the stars are still accreting gas, so their final masses are likely to increase by a few percent. The error bars correspond to the Poisson statistical noise.}
\label{fig:mspectrum}
\end{figure}

The mass distribution of the objects produced by disc fragmentation is shown in Fig.~\ref{fig:mspectrum}. $\,\sim\!70\%$ of the stars produced have masses below the H-burning limit, and $\,\sim\!3\%$ of these have masses below the D-burning limit. The mass distribution peaks around $30\;{\rm to}\;40\,{\rm M}_{\rm J}$. It decreases smoothly towards higher masses, roughly as $d{\cal N}/dM\propto M^{-\alpha}$, with $\alpha \simeq 1.4$. There is no significant discontinuity across the H-burning limit at $\sim 80\,{\rm M}_{\rm J}$. The mass distribution drops precipitously towards lower masses, reflecting that the minimum mass for star formation is $M_{_{\rm MIN}}\sim5\,{\rm M}_{\rm J}$. The small number of stars that end up below the D-burning limit are the ones that are ejected from the ur-disc by a 3-body interaction, almost as soon as they have formed.

%%%%%%%%%%%%%%
%%%%%%%%%%%%

We emphasise that the mass function presented in Fig.\ref{fig:mspectrum} represents stars formed by fragmentation of a particular ur-system (i.e. an ur-disc with mass $M_{_{\rm DISC}}=0.7\,{\rm M}_{\sun}$, radius $R_{_{\rm DISC}}=400\,{\rm AU}$, etc. in orbit round an ur-star with mass $M_{1}=0.7\,{\rm M}_{\sun}$). Therefore it can not be compared meaningfully with the overall stellar IMF. Rather it will be necessary first to repeat the numerical experiment reported here for many different ur-systems (i.e. different combinations of $M_{_{\rm DISC}}$, $R_{_{\rm DISC}}$, $M_{_1}$, etc.), and then to convolve the results with the appropriate distributions of $M_{_{\rm DISC}}$, $R_{_{\rm DISC}}$ and $M_{_1}$, before comparing with the observed stellar IMF. A Monte Carlo experiment addressing this aspect of the problem is in progress (Attwood et al., in prep.)

\subsection{Radial distribution and the brown dwarf desert}
\label{sec:rdistribution}

%%%%%%%%%%%%%%%
\begin{figure*}
\centerline{
\includegraphics[height=6.2cm]{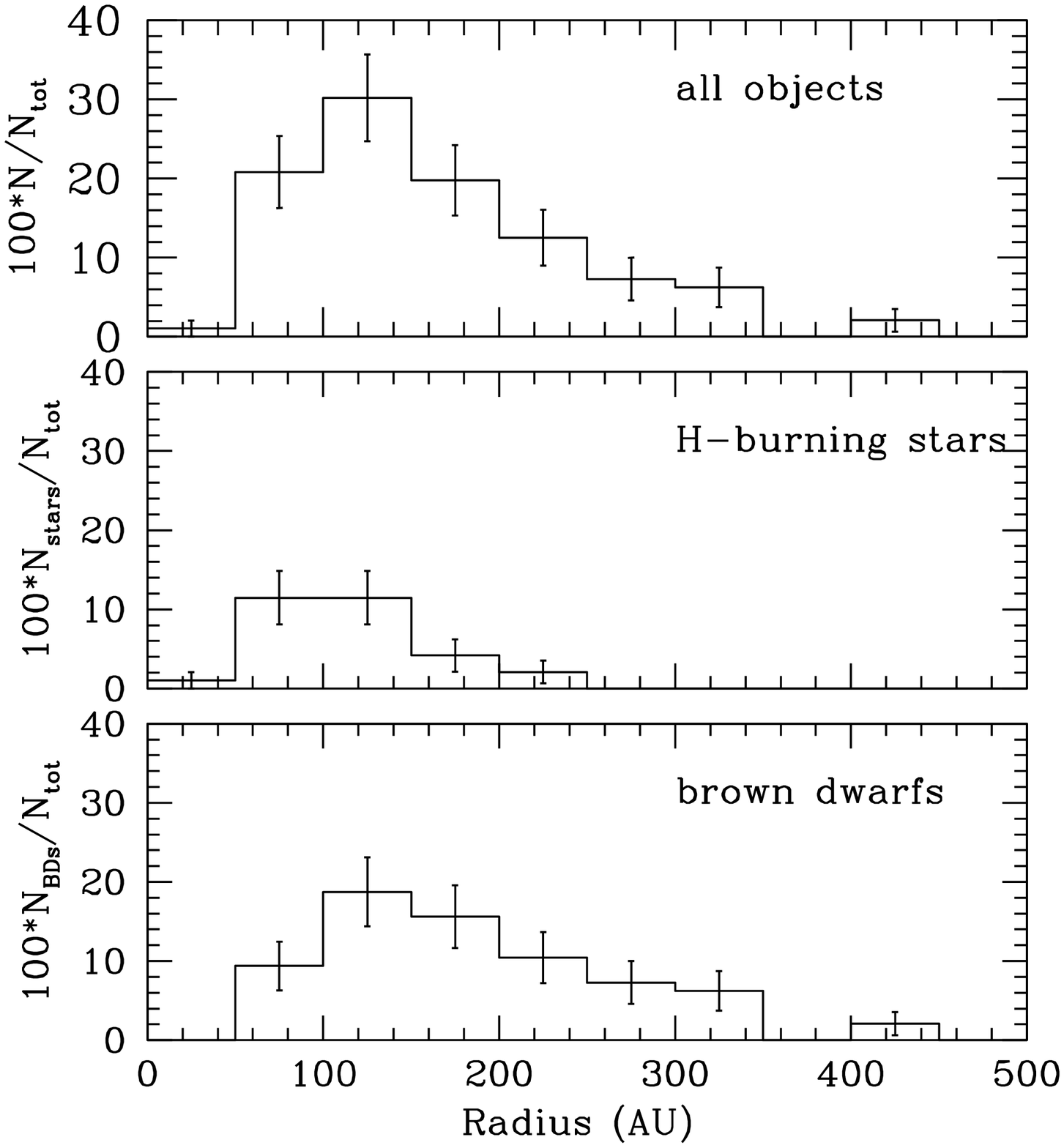}\hspace{-0.4cm}
\includegraphics[height=6.2cm]{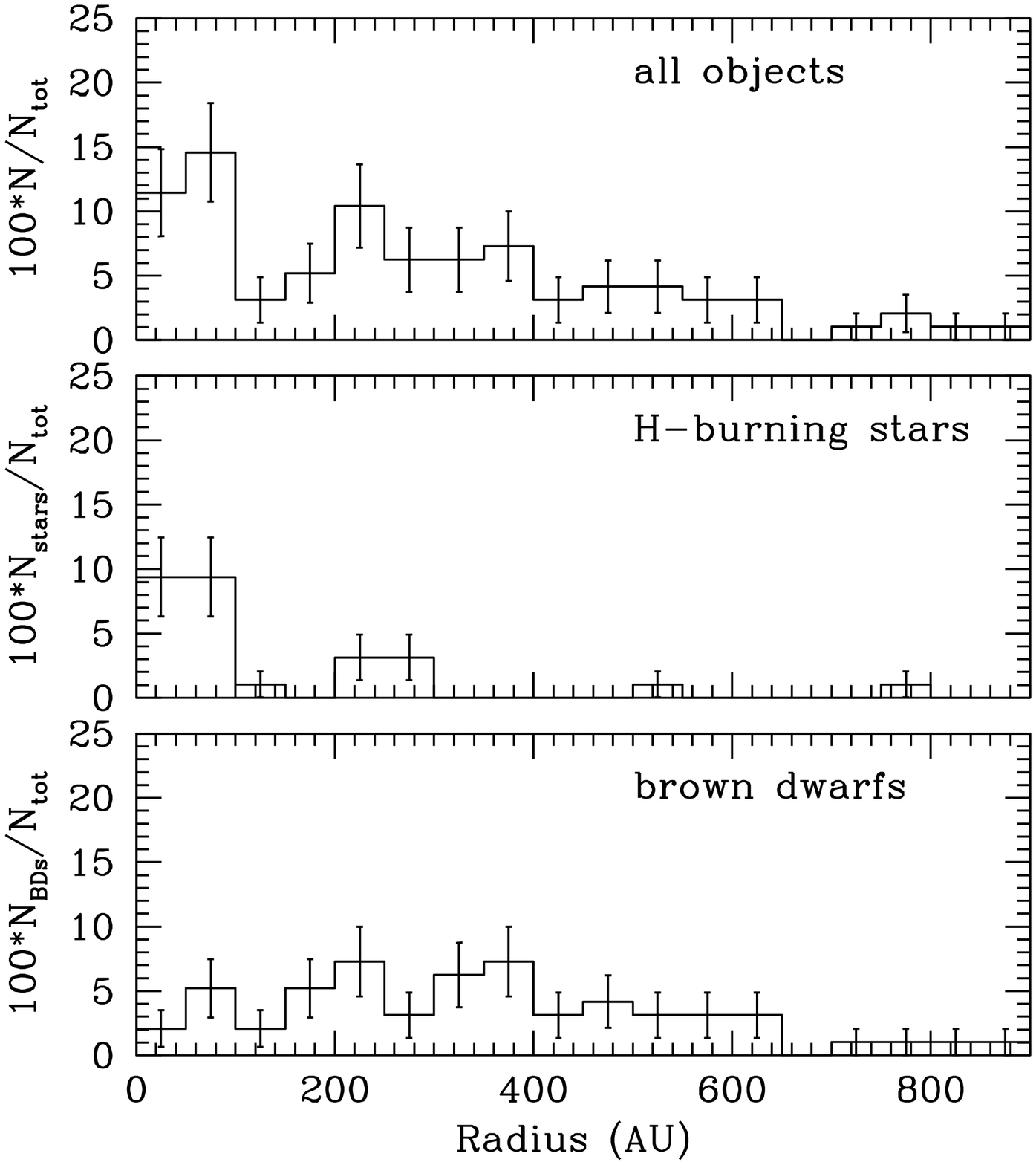}\hspace{-0.4cm}
\includegraphics[height=6.2cm]{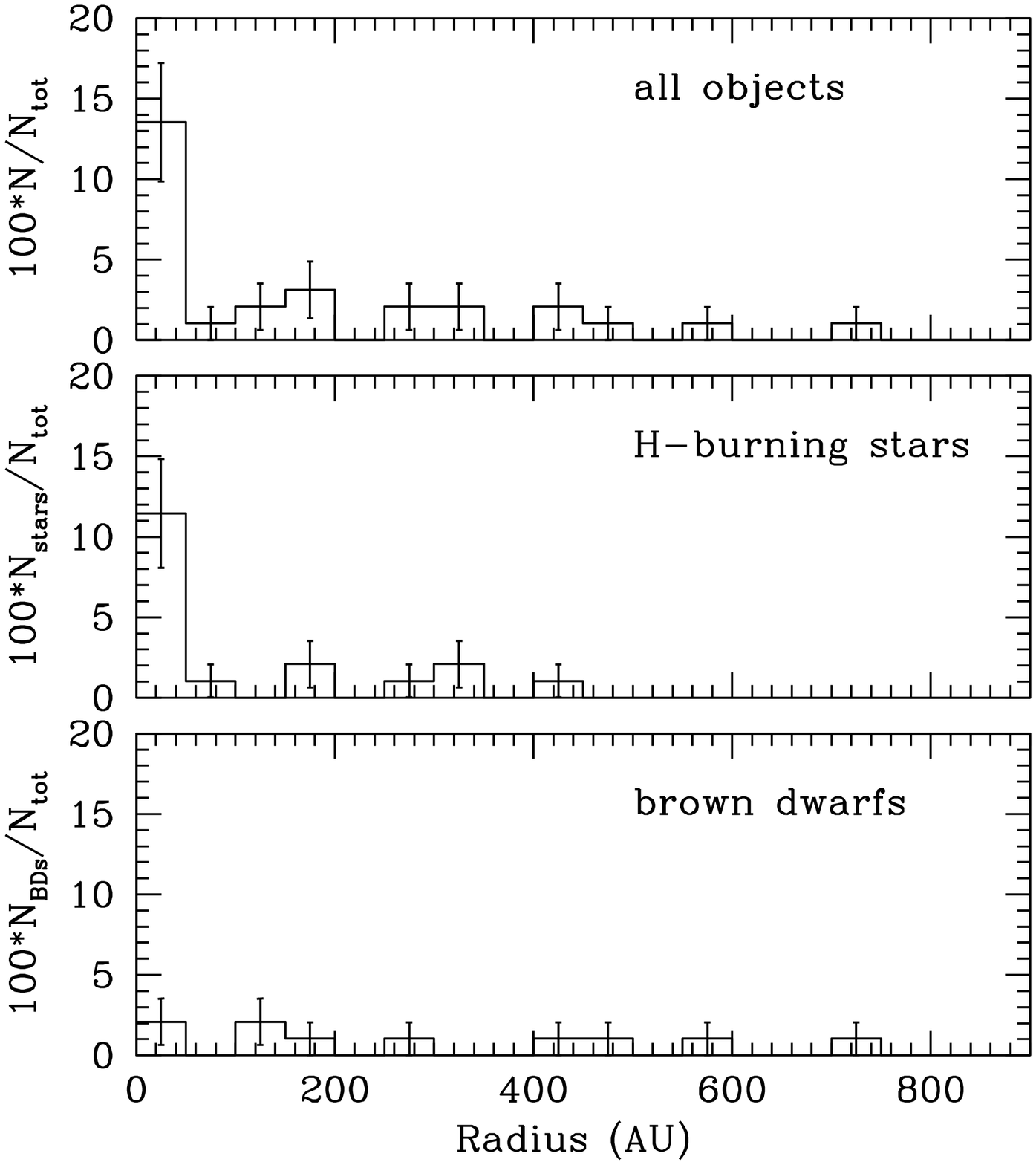}}
\caption{Radial distribution of the stars, (i) at the moment of formation ($\sim 5\,{\rm kyr}$; left), (ii) at the end time of the SPH simulation ($\sim 20 \,{\rm kyr}$; centre), and (iii) at the end of the N-body simulation ($200\,{\rm kyr}$). The radial distribution is plotted for all stars (top), for the HBs alone (middle), and for the BDs alone (bottom).}
\label{fig:rspectrum}
\end{figure*}
%%%%%%%%%%%%%

Fig.~\ref{fig:rspectrum} shows the radial distribution of the stars formed by disc fragmentation, i.e. the distance $R\;$ from the ur-star. Radial distributions are shown for all stars, and then separately for HBs and BDs, (i) at the moment of formation ($\sim 5 \,{\rm kyr}$; left), (ii) at the end of the SPH simulation ($\sim 20 \,{\rm kyr}$; centre), and (iii) at the end of the N-body simulation ($200 \,{\rm kyr}$; right). Stars are categorised as HBs or BDs on the basis of their final masses; here the BD category includes the PMs. 

The most likely location at which stars form is between $100\;{\rm and}\;200\,{\rm AU}$ from the ur-star (Fig.~\ref{fig:rspectrum}; left). This is the combined effect of the two criteria for disc fragmentation discussed in Section \ref{SEC:INTRO}.

Close to the central primary star ($R\la 60\,{\rm AU}$), there is ample mass for fragmentation, so the Toomre criterion is satisfied ($Q\leq 1$). However, the cooling time is high compared with the orbital period, and hence stars cannot condense out fast enough and they are quickly sheared apart. This minimum radius for fragmentation agrees well both with analytical predictions (Rafikov 2005; Whitworth \& Stamatellos 2006), and with previous numerical simulations (see also Stamatellos et al. 2007b, Stamatellos \& Whitworth 2008).

Conversely, far from the ur-star ($R>300\,{\rm AU}$), the cooling time is quite low, but there is only just enough mass for the disc to be Toomre unstable ($Q\sim 1$), and so fragmentation is sluggish.

At locations between $R\sim 100\;{\rm and}\;R\sim 200\,{\rm AU}$ the conditions are just right for vigorous fragmentation to occur.

In the lower panels of Fig.~\ref{fig:rspectrum} (left), we show separately the formation locations of HBs and BDs. Stars forming close to the ur-star generally accrete more gas and so their masses tend to increase above the H-burning limit.

The initial radial distribution quickly changes due to dynamical interactions between the newly-formed stars. In each disc between 5 and 11 stars form between 60 and 400 AU. Therefore the probability for interactions is high. In Fig.~\ref{fig:rspectrum} (centre) we present the radial distribution of the stars at the end of the SPH simulation at $\sim 20 \,{\rm kyr}$. The HBs are mostly within 100 AU of the ur-star, whereas the BDs are mostly outside 100 AU.

The radial distribution has changed even more by the end of the $N$-body evolution, after $200 \,{\rm kyr}$ (Fig.~\ref{fig:rspectrum}; right). Most HBs have remained bound to the ur-star (half at close distances, $\sim 10\,{\rm AU}$ and the rest on wider orbits of a few hundred AU). On the other hand, most of the BDs have been ejected, and those that remain bound to the ur-star have wide orbits of a few hundred AU.

\begin{figure}
\centerline{
\includegraphics[width=0.5\textwidth]{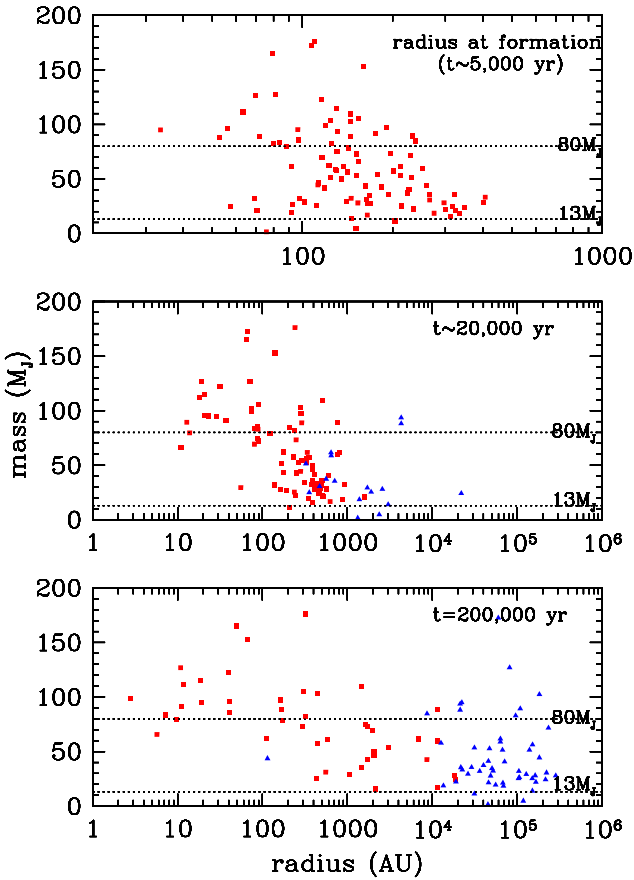}}
\caption{The origin of the brown dwarf desert. The formation radius (top), the radius at $t\sim 20 $ kyr (middle), and the radius at $t=200$ kyr, are plotted against the final stellar mass. HBs are scattered towards the ur-star, whereas BDs are ejected or migrate towards wider orbits. Hence, the region close to the ur-star is exclusively populated by low-mass HBs. The blue triangles correspond to objects that are ultimately not bound to the ur-star (ejected or about to be ejected from the system).}
\label{fig:rmasslg}
\end{figure}

This is also seen on Fig.~\ref{fig:rmasslg}, where the object final mass is plotted against the radius at formation (top), the radius at the end of the SPH simulation ($20 \,{\rm kyr}$; middle), and the radius at the end of the N-body simulation ($200 \,{\rm kyr}$; bottom).  Both HBs and BDs form over a wide range of radii, but after the system has evolved for 200 kyr there are 13 HBs orbiting within 60 AU of the ur-star, and only 1 BD. The HBs, being more massive, tend to form at smaller radii, {\it and} during dynamical 3-body interactions the more massive stars tend to become more tightly bound.

This phenomenology provides a natural explanation for the brown dwarf desert, i.e. the lack of BDs as close companions to Sun-like stars ($R\la 5\,{\rm AU}$; Marcy \& Butler 2000) in contrast with the relatively high frequency of both HBs and planets at these radii. First, stars with masses below the H-burning limit only condense out of the ur-disc at large radii ($\ga 100\,{\rm AU}$) -- as predicted analytically by Whitworth \& Stamatellos (2006). Second, such stars are only likely to stay below the H-burning limit if they remain at large radius or are ejected from the ur-disc. Third, the interactions which occur between stars formed in the ur-disc tend to scatter the more massive ones inwards and the less massive ones outwards. Consequently the more massive stars tend to end up in the inner parts of the ur-disc where there is still gas to be accreted, and hence they tend to grow even more massive until they exceed the H-burning limit.

At the end of our simulations, only one of the 12 ur-stars has a close ($<30$~AU) brown-dwarf companion, whereas there are 12 low-mass HBs in the same radial interval. Hence, it is $\sim 12$ times more likely for the close companion of a Sun-like star to be an HB, rather than a BD. This is comparable with observations, which suggest that $<0.5\%$ of Sun-like stars have BD companions within $\sim 5\,{\rm AU}$ (e.g.  Marcy \& Butler 2000; Udry et al. 2003), or $<2\%$  within  $\sim 8\,{\rm AU}$ (e.g in the Hyades cluster, Guenther et al. 2005). This fraction rises to $\sim 13\%$ for HB companions (Duquennoy \& Mayor 1991), i.e. HB companions are between 7 and 25 times more frequent than BD companions.

%%%%%%%%%%%%%%
%%%%%%%%%%%%

In fact, at the end of our simulations only 4 out of 24 ($\sim 17\%$) of the stars within $\sim 400\,{\rm AU}$ of the ur-star are BDs, suggesting that the brown dwarf desert extends to a wider region around Sun-like stars.  This is consistent with the results of the Gemini Deep Planet Survey that found a very low probability of BDs (with $m<40$~$M_{\rm J}$ in the region 25-250~AU ($\sim$~1.9\%; Lafreni{\`e}re et al. 2007). Hence, systems like 1RXS J160929.1-210524, a PM with mass $8^{+4}_{-1}$~M$_{\rm J}$ orbiting a star with $0.85^{+0.2}_{-0.1}$~M$_{\sun}$ at $\sim 330$~AU (Lafreni{\`e}re et al. 2008), can be explained with this model; however, such systems should be rare. 

In contrast, BDs dominate the region from $\sim 400$ to $\sim 10,000\,{\rm AU}$; 19 out of 22 ($\sim 86\%$) of the stars at these large radii are BDs. Gizis et al. (2001) report that at separations $\ga 1000\,{\rm AU}$ Sun-like stars have comparable numbers of brown dwarf and M dwarf companions. 

Once again, we should not necessarily expect a detailed correspondence with the observations, because we have only studied a single ur-system (i.e. a single combination of $M_{_{\rm DISC}}$, $R_{_{\rm DISC}}$, $M_{_1}$, etc.).

%%%%%%%%%%%%%%
\begin{figure}
\centerline{
\includegraphics[height=11cm]{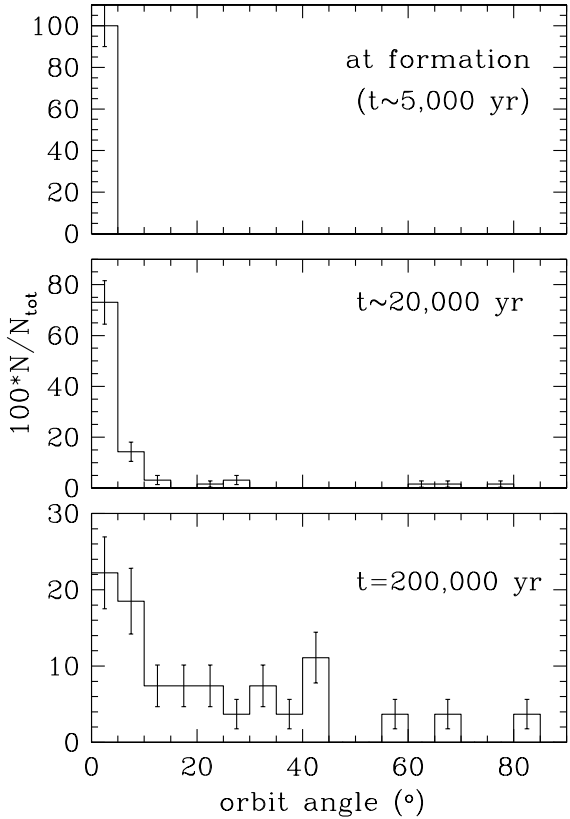}}
\caption{The distribution of inclination angles between the orbital planes of the stars and the orbital plane of the ur-disc, at formation ($t\sim 5 \,{\rm kyr}$; top), at $t\sim 20 $ kyr (middle) and at $t=200$ kyr (bottom). The stars form very close to the ur-disc midplane (within $\pm 4\degr$) but due to 3-body interactions many of them acquire inclined orbits. Coplanarity is not a characteristic of disc fragmentation.}
\label{fig:anglespec}
\end{figure}
%%%%%%%%%%%%

\subsection{Orbital plane of the bound objects}

Initially (at formation time, $\sim 5 \,{\rm kyr}$) the orbits of the stars formed in the ur-disc are almost coplanar with the ur-disc; their orbital planes all lie within $5\degr$ of the ur-disc plane (Fig.~\ref{fig:anglespec}, top). However, due to 3-body interactions the stars start to acquire inclined orbits,. By $\sim 20\,{\rm kyr}$ only $\sim 70\%$ of the orbital planes lie within $5\degr$ of the ur-disc plane, and by $\sim 200 \,{\rm yr}$ this percentage has dropped to $\sim 22\%$ (Fig.~\ref{fig:anglespec}, middle and bottom). We expect that in a realistic asymmetric disc, still accreting from its natal cloud, these percentages will be even smaller.

The inclination of the orbital plane does not appear to depend on the mass of the star, or its distance from the ur-star. Consequently, coplanarity in observed multiple systems, or the absence of it, cannot confirm or rule out formation by disc fragmentation.

%%%%%%%%%%%%%%
\begin{figure}
\centerline{ 
\includegraphics[height=7cm,angle=-90]{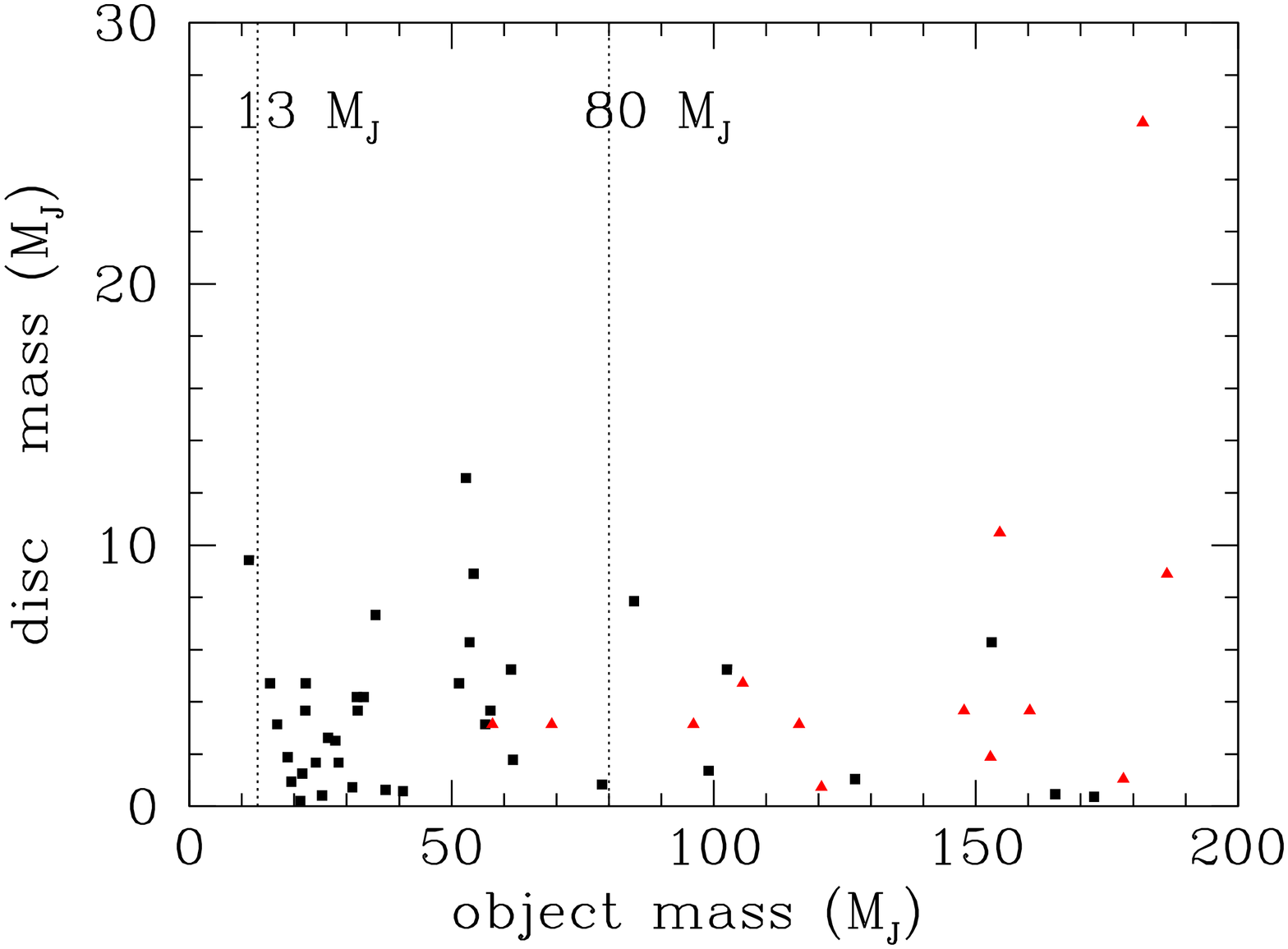}}
\centerline{
\includegraphics[height=7cm,angle=-90]{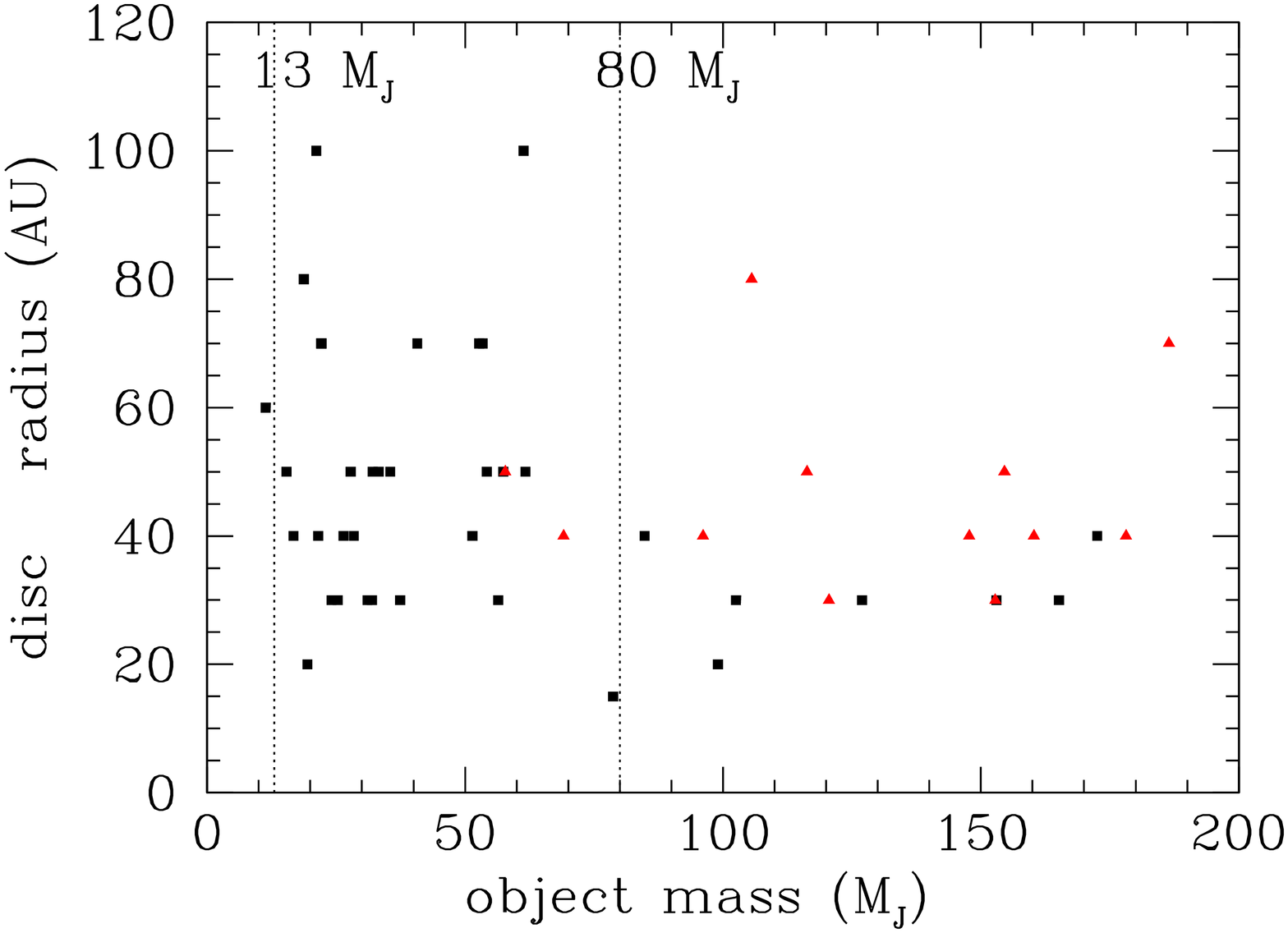}}
\caption{Disc masses (top) and radii (bottom) plotted against the masses of the host stars. The red triangles correspond to circumbinary discs.}
\label{fig:mrdiscmstar}
\end{figure}
%%%%%%%%%%%%

%%%%%%%%%%%%%%
\begin{figure}
\centerline{
\includegraphics[height=7 cm,angle=-90]{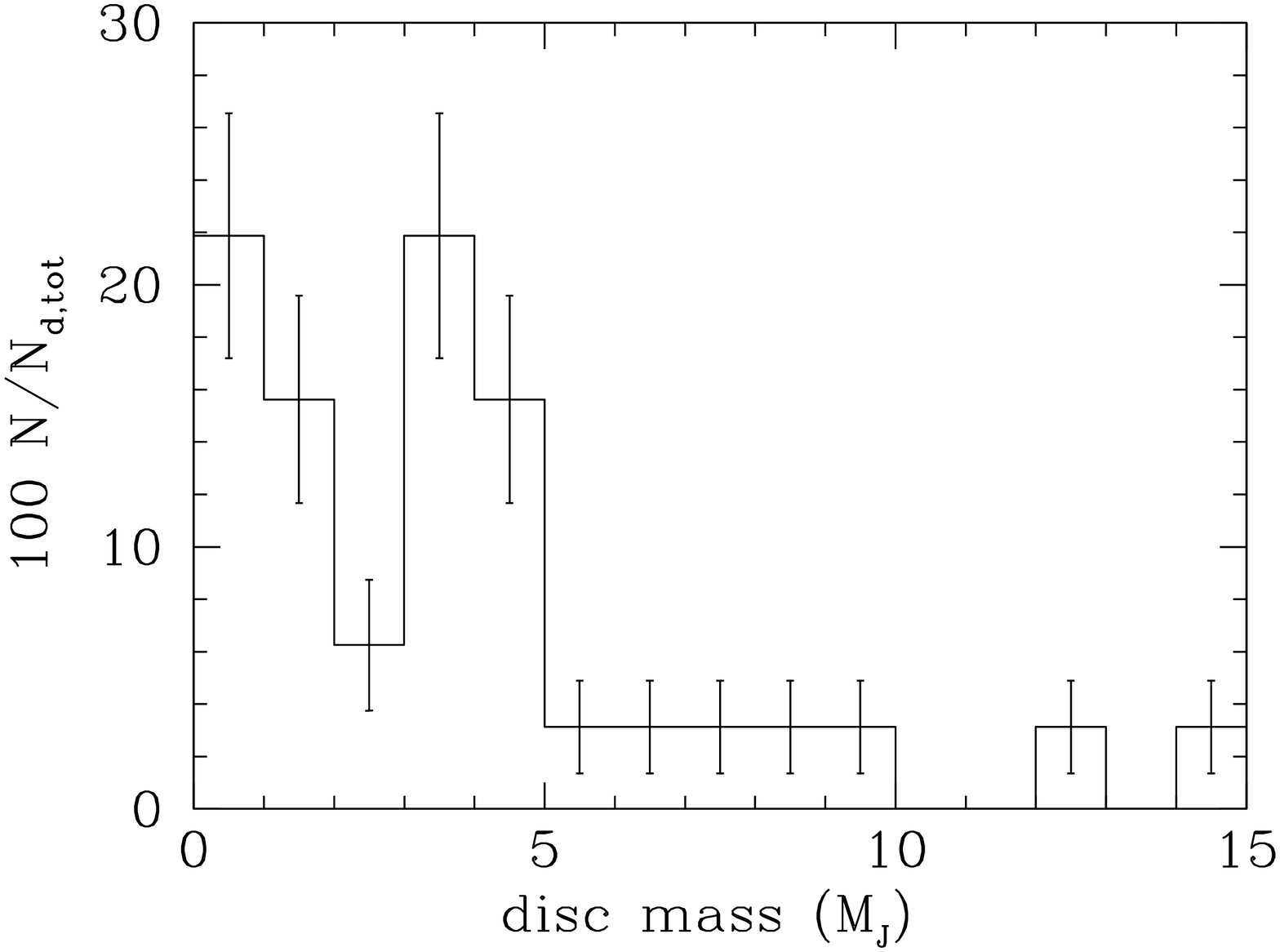}}
\centerline{
\includegraphics[height=7 cm,angle=-90]{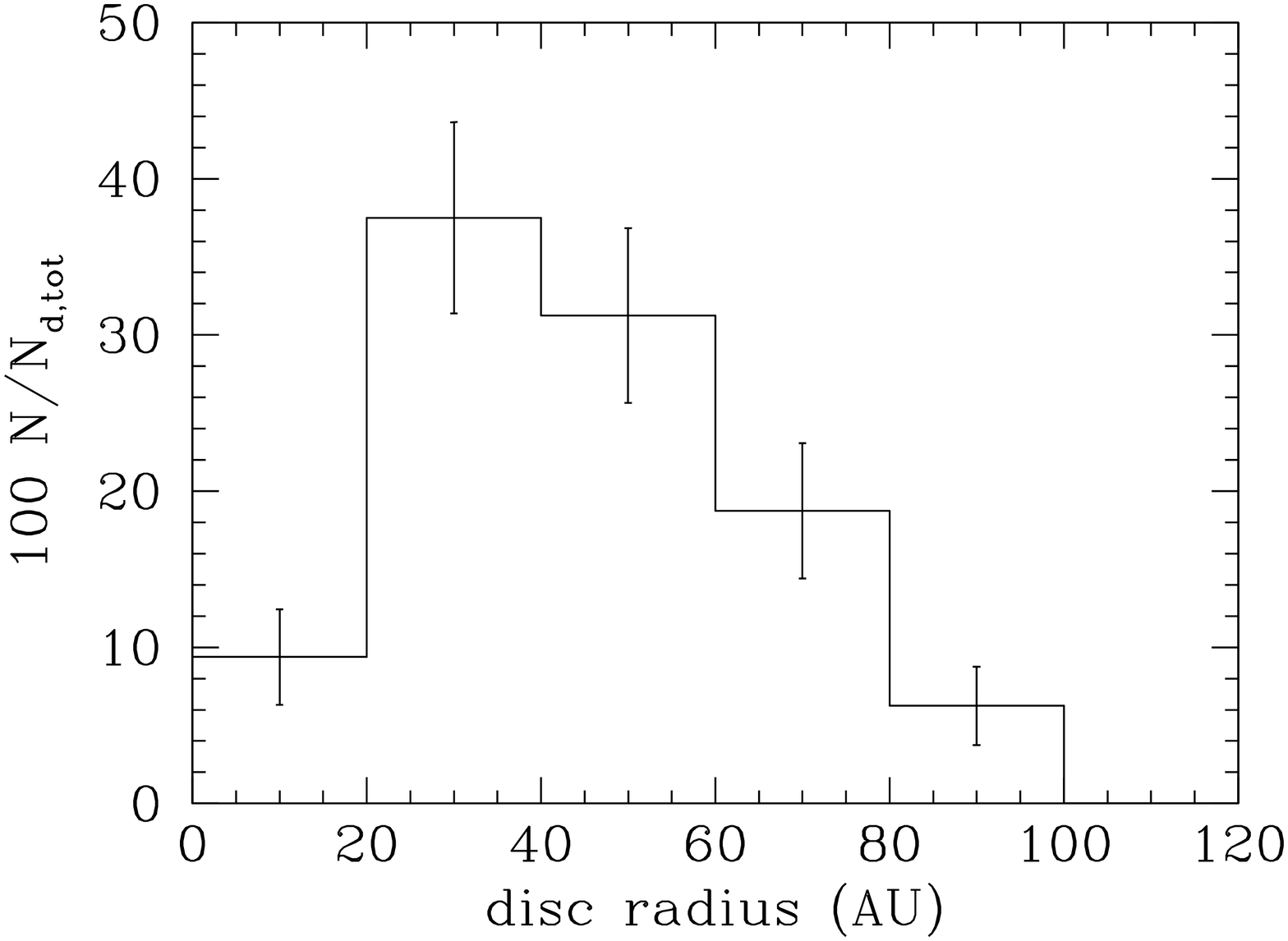}}
\caption{Mass (top) and size (bottom) distribution of the discs around brown dwarfs formed by  fragmentation of the ur-disc.}
\label{fig:mrdiscspecBD}
\end{figure}
%%%%%%%%%%%%

\subsection{Disc properties of low-mass objects formed by disc fragmentation}

When stars condense out of the ur-disc, they are normally attended by their own individual accretion discs, but some of these individual discs are subsequently stripped by dynamical interactions. At the end of the SPH simulation ($\sim 20 \,{\rm kyr}$), $\sim 70\%$ of the stars still have individual accretion discs. In addition, some of the binary systems have circumbinary discs.

These percentages are comparable with, but somewhat higher than, those reported in the literature. Jayawardhana et al. (2003) found a disc fraction of $40\;{\rm to}\;60\%$ in a mixed sample from Chamaeleon I, IC 348, Taurus, and U Sco, using $JHKL'$ photometry. Luhman et al. (2005c), using Spitzer NIR observations, found a disc fraction of $42\pm13\%$ for IC348 and $45\pm7\%$ for Chamaeleon I. However, these observed statistics correspond to stars in the fields of these clusters (i.e. stars which, if they have been formed by fragmentation of an ur-disc, have since been ejected) and so we expect their disc frequency to be somewhat reduced. 

Of the 13 systems (11 singles and 2 binaries) ejected by the end of our SPH simulations at $\sim 20 \,{\rm kyr}$, only one of the singles, but both the binaries, retain discs. Thereafter, the simulation switches to an $N$-body code, and so we cannot say anything quantitative about how the disc statistics subsequently evolve. However, since there are many stars with discs orbiting at very large distances from the ur-star -- for example,  at the end of the SPH simulations, there are 16 systems, including 2 binaries, orbiting at $R\ga 500\,{\rm AU}$ -- we should expect many of these to be liberated from the ur-star by the tides of passing stars, and to retain their discs (e.g. Whitworth \& Stamatellos 2006; Goodwin \& Whitworth 2007). Thus we infer that young BDs in the field should have a significant disc fraction, and that young BDs orbiting Sun-like stars at large radius should have an even higher disc fraction.

The masses and radii of individual discs are plotted against the masses of their host stars on Fig.~\ref{fig:mrdiscmstar}. The disc masses are generally $\la 10\,{\rm M}_{\rm J}$, but there are a couple of discs with higher masses (a few tens of M$_{\rm J}$). There seems to be no correlation between the disc masses and the star masses. The disc radii range up to $100\,{\rm AU}$, but most of them are $\la 70\,{\rm AU}$. Again there seems to be no correlation between disc radius and star mass.

The individual discs around BDs and HBs are statistically indistinguishable. The distributions of mass and radius for discs around BDs are shown in Fig.~\ref{fig:mrdiscspecBD}. Most BD discs ($\sim 80\%$) have masses below $\sim 5\,{\rm M}_{\rm J}$, but the mass can be as high as $\sim 15\,{\rm M}_{\rm J}$ (Fig.~\ref{fig:mrdiscspecBD}, top). Most BD discs have radii $\la 60\,{\rm AU}$, but the radius can be as high as $100\,{\rm AU}$ (Fig.~\ref{fig:mrdiscspecBD}, bottom). These disc masses and radii are consistent with observations, which show discs to be very common around BDs (Klein et al. 2003; Luhman 2004; Luhman et al. 2005a; Scholz et al. 2006; Guieu et al. 2007; Riaz \& Gizis 2007, 2008), even around very low-mass ones (e.g. Luhman et al. 2005b).  The BD discs  are massive enough to make the formation of rocky planets with masses up to 5~M$_{_\oplus}$ around BDs a plausible scenario, at least around some of them (Payne \& Lodato 2007).

%%%%%%%%%%%%%%
\begin{table}
\begin{minipage}{0.45\textwidth}
\caption{Properties of low-mass binaries (i.e. those in which both components have condensed out of the disc): $m_p$, the primary mass; $m_s$, the secondary mass; $q=m_s/m_p$, the mass ratio; $a_{_{\rm BIN}}$, the semi-major axis of the low-mass binary orbit; $e_{_{\rm BIN}}$, the eccentricity of the low-mass binary orbit; $a$, the semi-major axis of the orbit of the low-mass binary around the ur-star (in those cases where the low-mass binary remains bound to the ur-star, otherwise blank). The binaries are grouped into four categories, according as the components are (i) both HBs, (ii) an HB and a BD, (iii) both BDs, and (iv) a BD and a PM.}
\label{tab:binaries}
\centering
\renewcommand{\footnoterule}{}  % to avoid a line before footnotes
\begin{tabular}{@{}cccccc} \hline
\noalign{\smallskip}
   $m_{\rm p}/$M$_{\rm J}$
&  $m_{\rm s}/$M$_{\rm J}$
& q 
& $a_{\rm bin}$/AU  & $e_{\rm bin}$ 
& $a$/AU \\
\noalign{\smallskip}
\hline
\noalign{\smallskip}
     96  & 86      & 0.90 &  0.3  & 0.4 & 20             \\
     98 &  89     & 0.91 &  1.4   & 0.5 & 140     \\
     176 &  82  &  0.47 &  1.5   & 0.8 & 230              \\
     88 &   83    &  0.94  &   0.3& 0.7 & -           \\
 \hline
 89    &  59     & 0.66 &  1.0    & 0.9 &    7700         \\
 102  & 44       & 0.43 &  0.3   & 0.8 &    -             \\
 109  &  35       & 0.33 &  0.6  & 0.8 &    800               \\
 105  &   73    & 0.70 &  0.5    &  0.3 &    200              \\
 \hline
  50    &  46     &  0.92   &     0.6  & 0.5 & 1500              \\
  28   &   24   &  0.86     &    235  & 0.7  & 4000              \\
  62   &   59    &  0.95    &    0.6  & 0.9 & -              \\
  73    &   43    &  0.59   &     1.3 & 0.7 & 1350              \\
  \hline
   53   &  11      &  0.21  &    112  & 0.6 & -              \\
\noalign{\smallskip}
\hline
\end{tabular}
\end{minipage}
\end{table}
%%%%%%%%%%%

\subsection{Low-mass binary properties}

Here we discuss the statistical properties of the binary systems that exist at the end of the simulations. We confine the discussion to binaries in which both components are stars that have condensed out of the disc, and we refer to these as low-mass binaries. There are also binary systems and higher multiples in which one of the components is the ur-star; we do not discuss these here. 

Our simulations produce 13 low-mass binaries, of which 4 comprise two low-mass HBs, 4 comprise a low-mass HB and a BD, 4 comprise two BDs, and one comprises a BD and a PM. The main properties of these binaries are recorded in Table~\ref{tab:binaries}. In total, $27\%$ of the stars that form end up in binary systems, corresponding to a binary fraction of $16\%$. This is comparable with the low-mass binary fraction in Taurus-Auriga ($\la 20\%$, Kraus et al. 2006), Chamaeleon I ($11^{+9}_{-6}\%$, Ahmic et al. 2007), and the field (e.g. $15\pm5 \%$, Gizis et al. 2003). However, these are optical surveys and they are not able to detect tight binaries with separation of a few AU. If we consider only the 2 wide binaries then the binary fraction we estimate is $\sim2.5\%$, which is smaller than observed. Very tight binaries can be probed by  radial velocity surveys (Joergens 2006a; Kurosawa et al. 2006; Maxted et al. 2008; Joergens 2008). These surveys find  a low-mass tight-binary fraction of $10-30\%$ which is consistent with our model.

%%%%%%%%%%%%%%%%%%%%%%%%%%%%%%%%%
\begin{figure}
\centerline{
\includegraphics[height=0.45\textwidth]{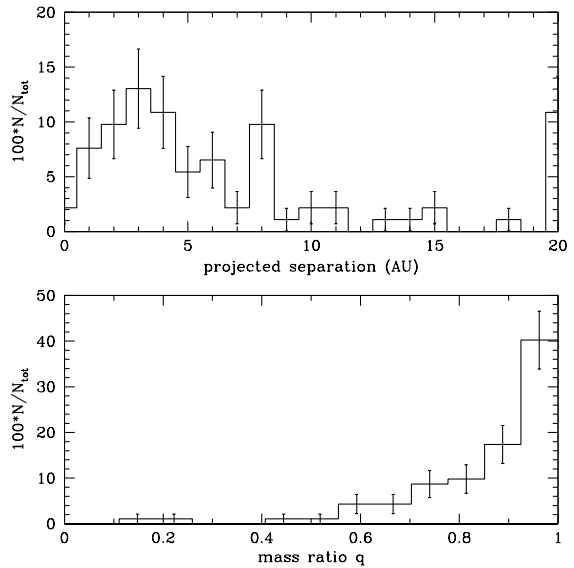}}
\caption{The observed properties of very low-mass binaries: projected separation (top) and mass ratio (bottom) distributions (data are taken from {\tt http://www.vlmbinaries.org}; last updated on 4 February 2008).}
\label{fig:vlmbinaries}
\end{figure}
%%%%%%%%%%%%%%%%%%%%%%%%%%%%%%%%%

11 of the 13 low-mass binaries have semi-major axes $a_{_{\rm BIN}}<2\,{\rm AU}$. Thus we predict that close low-mass binaries should outnumber wide ones, and this seems to be what is observed (e.g. Burgasser et al. 2007). In Fig.~\ref{fig:vlmbinaries} (top) we plot the projected separation distribution for the very low-mass binaries from the  {\tt http://www.vlmbinaries.org} dataset. In this dataset only $\sim 10\%$ of the binaries have separations $>20$~AU, which is similar to the predictions of our model (2/13 binaries, i.e. 15\%). We note that in Table~\ref{tab:binaries} we quote the semi-major axis of the binary, whereas Fig.~\ref{fig:vlmbinaries} refers to the projected separation, which for eccentric binaries is most often larger than semi-major axis). The two wide low-mass binaries -- with $a_{_{\rm BIN}}=112\;{\rm and}\;235\,{\rm AU}$ -- are the systems with the lowest total mass.

4 low-mass binaries, including one of the very wide systems, have become unbound from the ur-star. Therefore the ejection mechanism does not militate against delivering wide low-mass binaries to the field. One of the unbound low-mass binaries comprises a BD and a PM; it is therefore quite similar to 2MASS 1207-3932 (Chauvin et al. 2004, 2005). 

Amongst the 9 low-mass binaries that remain bound to the ur-star, the semi-major axis $a$ of the orbit around the ur-star tends to increase as the total mass of the low-mass binary decreases. Thus most of the low-mass binaries with an HB primary orbit the ur-star within $1000\,{\rm AU}$, whereas most of the low-mass binaries with a BD primary orbit the ur-star outside $1000\,{\rm AU}$. 

All of the low-mass binaries have high eccentricities, as a result of the dynamical interactions which form them, and/or subsequent dynamical interactions with other stars that have condensed out of the ur-disc. The high eccentricities of the wide binaries should persist, but the high eccentricities of the closer binaries may be damped somewhat by tidal interaction between their attendant discs; these tidal interactions are not accounted for in our N-body simulations.

Most of the low-mass binaries ($55\%$) have components with similar masses ($q>0.7$), in agreement with the observed properties of low-mass binaries (e.g. Burgasser et al. 2007; see Fig.~\ref{fig:vlmbinaries}).

25 (out of 67, i.e. 37\%) of the BDs formed in the ur-disc remain bound to the ur-star. The remaining 42 (out of 67, i.e. 63\%) are ejected. Of the 25 BDs that remain bound to the ur-star, 10 ($40\%$) are in low-mass binaries. In contrast, of the 42 BDs that become unbound from the ur-star, only 5 ($12\%$) are in low-mass binaries (one with an HB primary). Hence, BDs that are companions to Sun-like stars are more likely to be in binaries (binary frequency 25\%) than BDs in the field (binary frequency $5\;{\rm to}\;8\%$). This trend is comparable to what is observed, although the observed binary frequencies are somewhat higher. Burgasser et al. (2005) report a binary fraction of $45^{+15}_{-13}\%$ for BD companions to Sun-like stars, and a binary fraction of only $18^{+7}_{-4}\%$ for BDs in the field.

%%%%%%%%%%%%%%
\begin{figure}
\centerline{
\includegraphics[height=11cm]{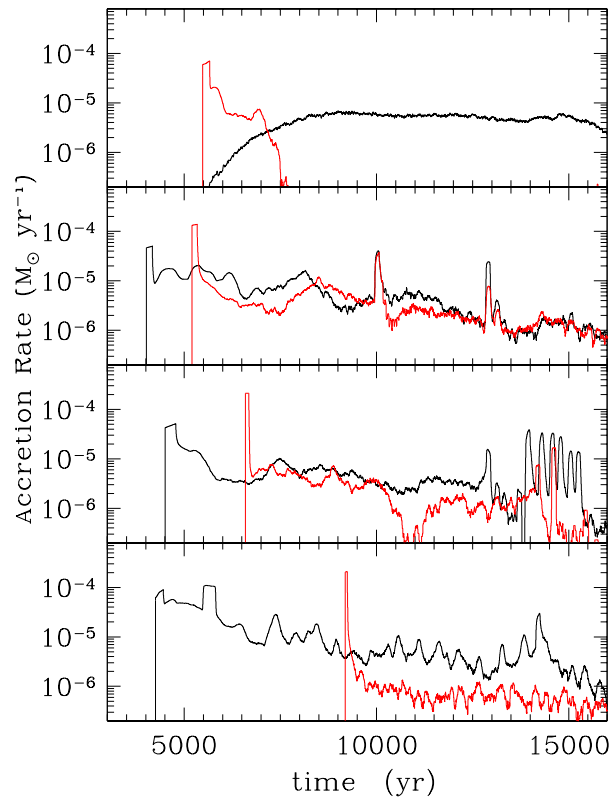}}
\caption{Accretion rates onto the stars forming in the simulation presented in Fig.~\ref{fig:rdm1}. The first plot shows the accretion rate onto the ur-star (black line) and the accretion rate onto a star which is ejected (red line). The second plot shows the accretion rates onto two stars which eventually become a binary with a circumbinary disc. Stars which form earlier and/or closer to the ur-star (black lines on the last 3 plots) tend to accrete more than stars which form later and/or further away from the ur-star (red lines on the last three plots). The accretion rates shows signs of periodic modulation (due to passing close to the ur-star and/or accretion from a circumbinary disc) and non-periodic modulation (when a star passes through a spiral arm or another star's disc).}
\label{fig:rdm1.accretion}
\end{figure}

\subsection{Accretion}
 
Newly formed stars continue to accrete material from the ur-disc. Stars forming closer to the ur-star tend to accrete more material than stars forming further away, firstly because there is more material in the inner parts of the ur-disc to accrete (particularly in the innermost part of the ur-disc, which is unable to fragment), and secondly because the more massive stars tend to migrate inwards. In our simulations, typical accretion rates are $\;\ga 10^{-6}\,{\rm M}_{\sun}\,{\rm yr}^{-1}$, which is high compared with the observed accretion rates onto brown dwarfs (typically $\;\la10^{-8}{\rm M}_{\sun}$; e.g. Natta et al.  2004; Muzerolle et al. 2005; Mohanty et al. 2005; Herczeg \& Hillenbrand 2008). However, the high accretion rates we record refer to an early, short-lived phase ($\la 15 \,{\rm kyr}$), and they are expected to decline significantly at later times.

The accretion rates often exhibit periodic modulations, during which the rate may increase by more than an order of magnitude. These modulations are partly due to high orbital eccentricity, and correspond to periastron passages within $R\sim 20\;{\rm to}\;50\,{\rm AU}$, occuring every few thousand years. There are also smaller modulations involving the components of close binaries as material is accreted onto them from a circumbinary disc. Finally, there are occasional episodic events when the accretion rate increases by up to an order of magnitude as a star passes through a spiral arm in the ur-disc or through the disc of another star.

In Fig.~\ref{fig:rdm1.accretion} we plot -- for the simulation presented in Fig~\ref{fig:rdm1} -- the accretion rates onto the ur-star and onto the 7 stars formed from the ur-disc. These plots are representative of the accretion rates onto stars in the other simulations. On the first plot, we present the accretion rate onto the ur-star (black line). The accretion rate initially increases, as the inner gap is filled in, and then remains approximately constant around $\dot{M}_{_1}\sim 5\times 10^{-5}$~M$_{\sun}\,{\rm yr}^{-1}$. The other star presented on the first plot (red line) is one which is ejected and so accretion is effectively terminated. There is actually a very low persistent accretion rate onto this star, because it retains a small accretion disc, but the rate is well below $10^{-8}\,{\rm M}_\odot\,{\rm yr}^{-1}$.

On the second plot we present the accretion rates onto two stars which form at different times but eventually become a binary system with a circumbinary disc. The star which forms later, and at larger distance from the ur-star, initially accretes less material than the star which forms earlier, and closer to the ur-star. After the binary forms, the accretion rates onto the two stars become very similar.

On the last two plots, the accretion rates onto the other four stars are presented. Again it can be seen that the stars which form later tend to accrete less than the stars which form earlier. In some instances the difference exceeds an order of magnitude. The stars which experience persistently higher accrertion rates tend to be the ones that form, and/or end up, closer to the ur-star.

\section{Aftermath}

\subsection{The ur-star and its companions}
 
%%%%%%%%%%%%%%
\begin{figure}
\centerline{
\includegraphics[height=8cm,angle=-90]{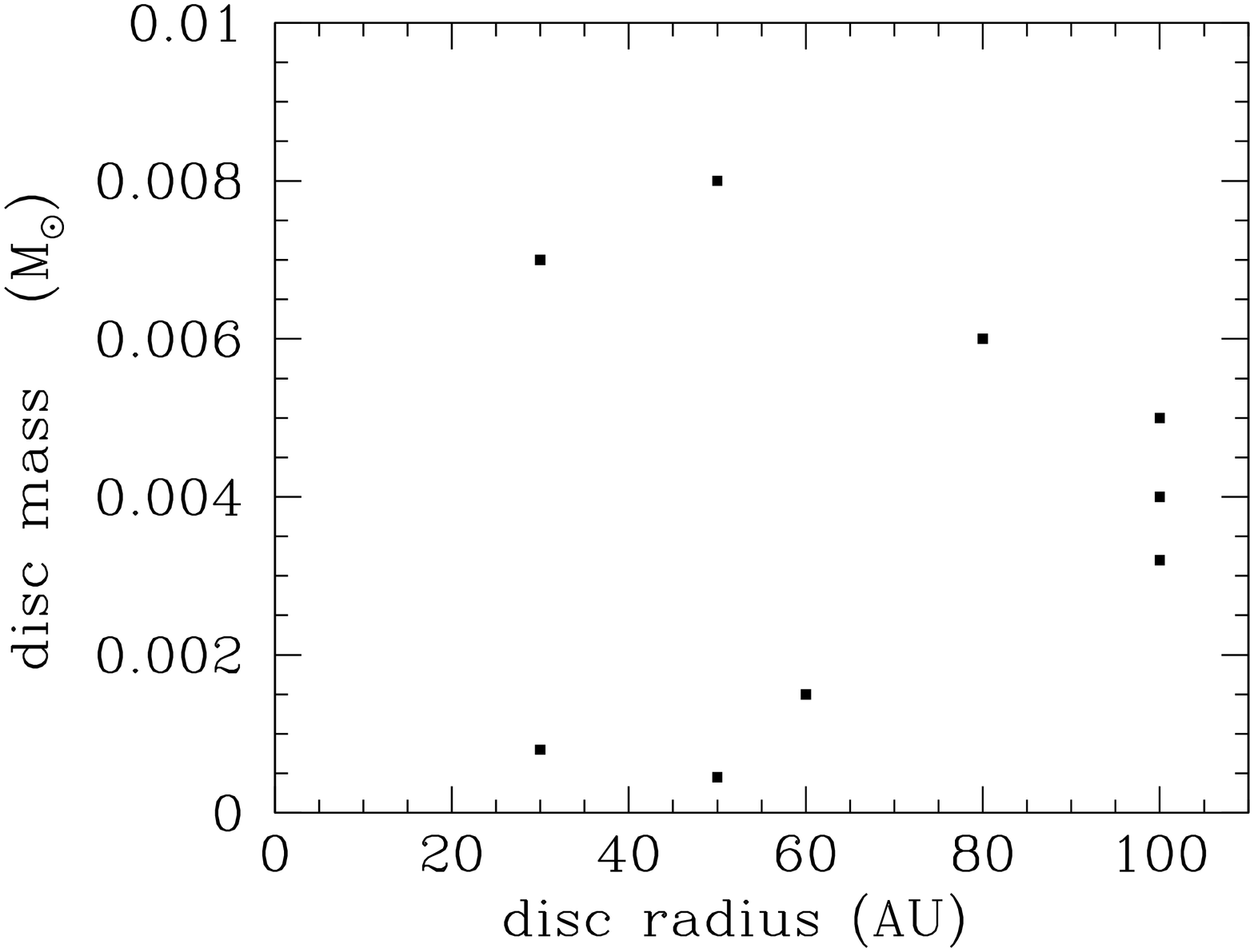}}
\caption{The masses and radii of the remnant ur-discs at the end of the SPH simulations, i.e. $t=20 \,{\rm kyr}$. Two cases fall outside this plot; they have $(M_{_{\rm DISC}},\,R_{_{\rm DISC}})\simeq(0.03{\rm M}_{\sun},\,200\,{\rm AU})$ and $(0.19\,{\rm M}_{\sun},\,150\,{\rm AU})$.}
\label{fig:endstardisc}
\centerline{
\includegraphics[height=8cm,angle=-90]{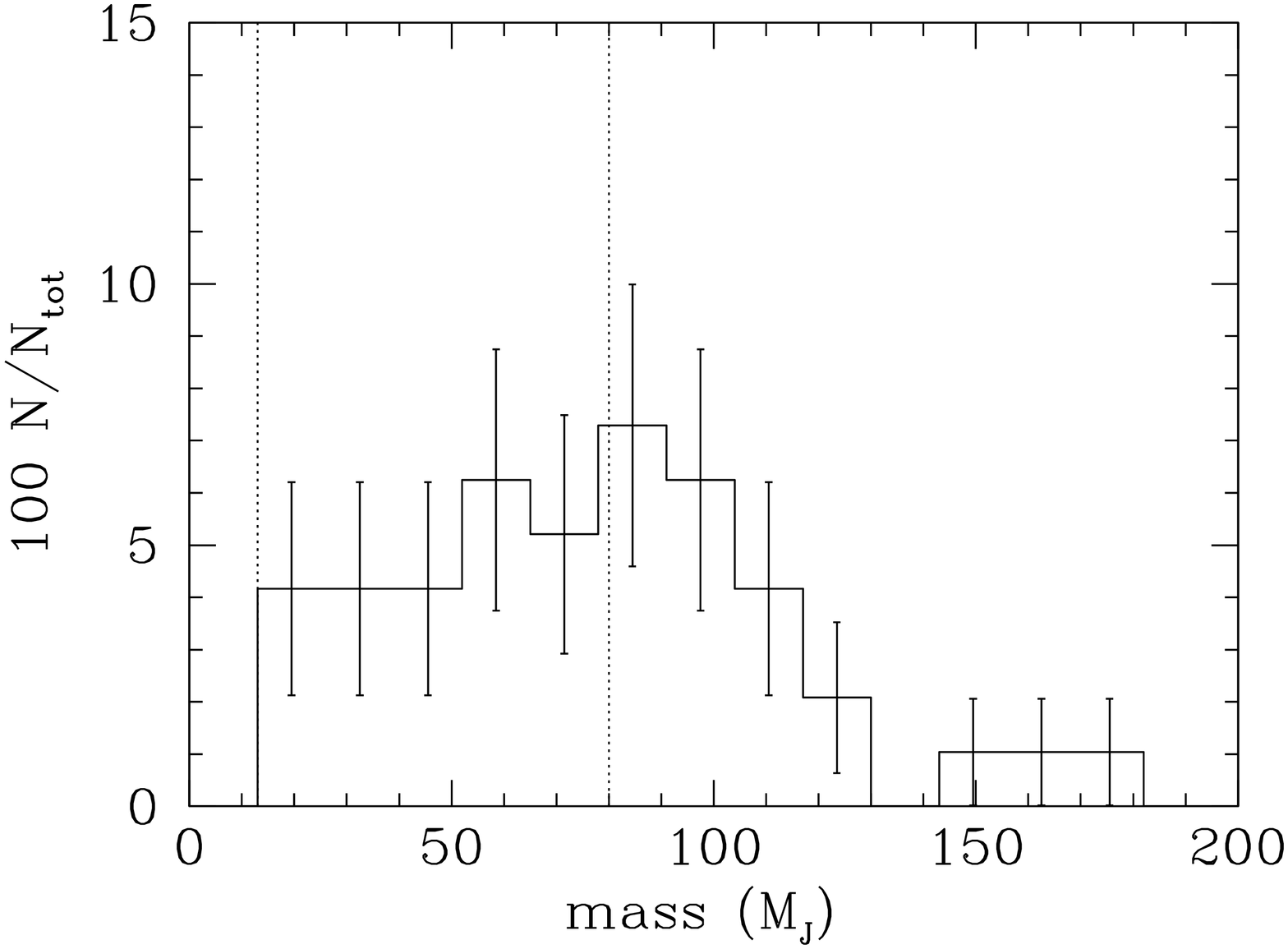}}
\caption{The mass distribution of stars which remain bound to the ur-star. There are almost equal numbers of HBs and BDs.}
\label{fig:mspecbound}
\end{figure}

\begin{figure}
\centerline{
\includegraphics[height=8cm,angle=-90]{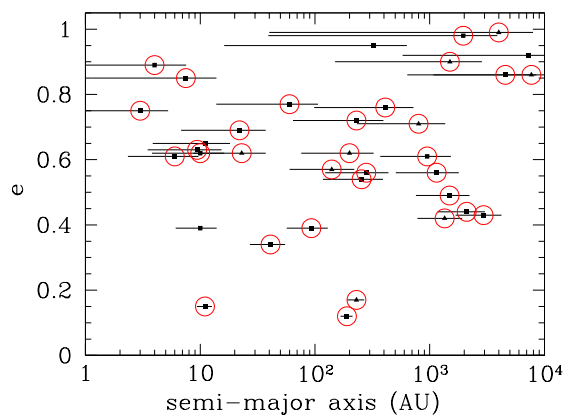}}
\caption{The orbital eccentricities and semi-major axes of stars which remain bound to the ur-star. The bars indicate the minimum and maximum extent of the orbit. The orbits are generally highly eccentric, as a result of dynamical interactions. The circles mark stars with discs, and the triangles correspond to binary systems.}
\label{fig:nbfinal.ecc}
\end{figure}
%%%%%%%%%%%%

The ur-discs dissipate on a dynamical timescale, i.e. within a few thousand years. In Fig.~\ref{fig:endstardisc} the mass of the remnant ur-disc is plotted against its radius. By this stage, most of the ur-discs have masses $M_{_{\rm DISC}}\la 0.01\,{\rm M}_\odot$, although one has $M_{_{\rm DISC}}\sim 0.2\,{\rm M}_\odot$. The ur-disc radii are all $R_{_{\rm DISC}}\la 200\,{\rm AU}$. These ur-disc masses and radii are typical of the discs observed around Sun-like stars (e.g. Mundy et al. 1995; Bally et al. 1998; Eisner et al. 2005, 2008; Williams et al. 2005; Eisner \& Carpenter 2006).

In our simulations the ur-star ends up with one or two H-burning companions and one or two brown-dwarf companions. The masses of the companions are almost equally distributed across the low-mass regime (see Fig.~\ref{fig:mspecbound}). One of the companions is in a relativel close orbit ($R\la 20\,{\rm AU}$), and this one is almost always an H-burning star. The other two or three companions orbit much further out ($R\ga 100\,{\rm AU}$), and often two of them are in a binary system. 

The eccentricities of these wide companions are high, due to dynamical interactions with other stars (Fig.~\ref{fig:nbfinal.ecc}). Thus, although their semi-major axes are greater than $1,000\,{\rm AU}$, due to their high eccentricity they pass within $\sim 100\,{\rm AU}$ of the ur-star with a periodicity between $10^4\;{\rm and}\;10^7\,{\rm yr}$, disrupting the remaining ur-disc and/or the orbits of any planets formed from the inner ur-disc (see Sect.~\ref{sec:planets}).

%%%%%%%%%%%%%%
\begin{figure}
\centerline{
\includegraphics[height=7.8cm,angle=-90]{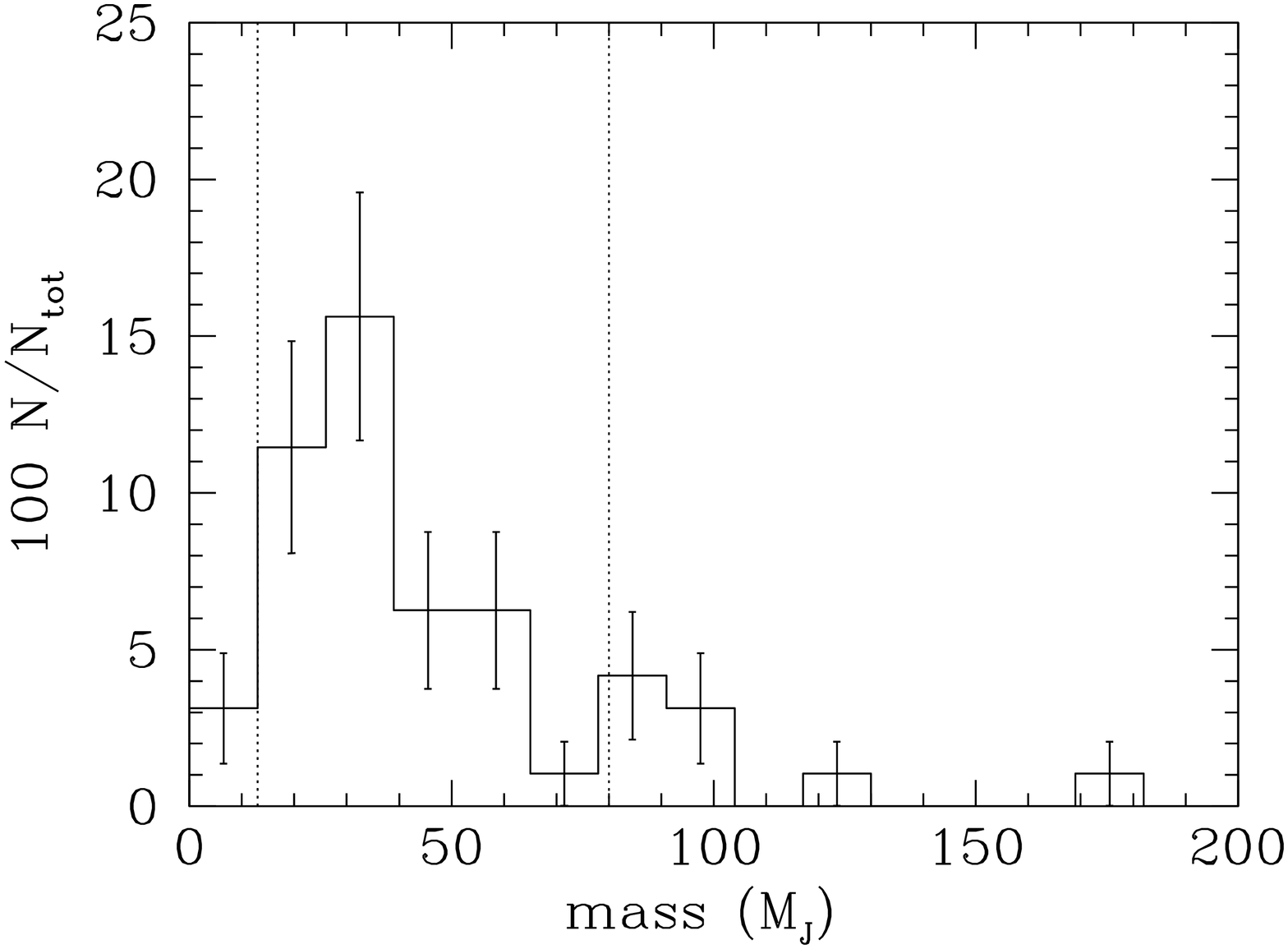}}
\caption{The mass distribution of the ejected stars. Most of them are low-mass BDs.}
\label{fig:mspec.ejected}
\centerline{
\includegraphics[height=8cm,angle=-90]{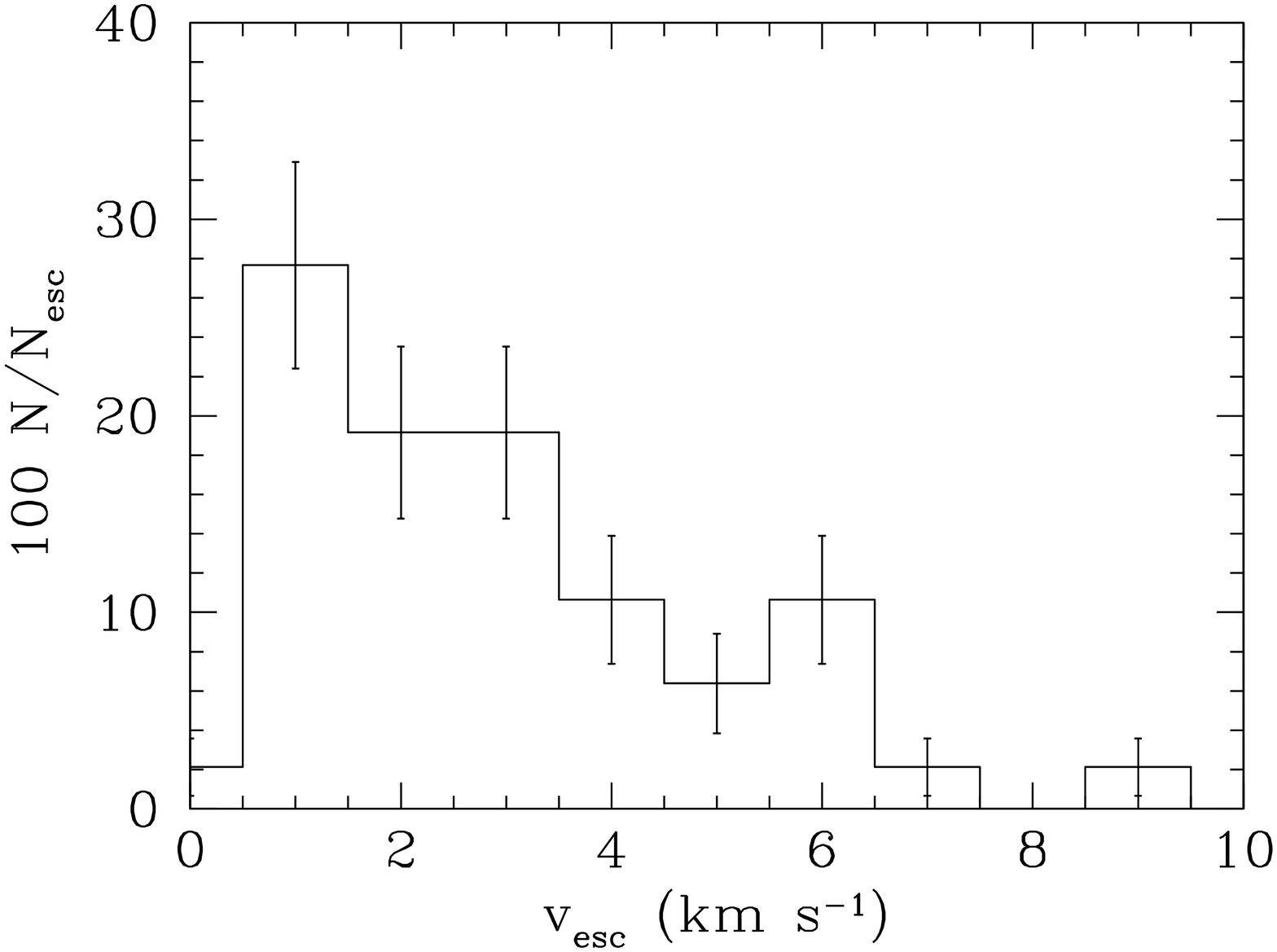}}
\centerline{
\includegraphics[height=7.5cm,angle=-90]{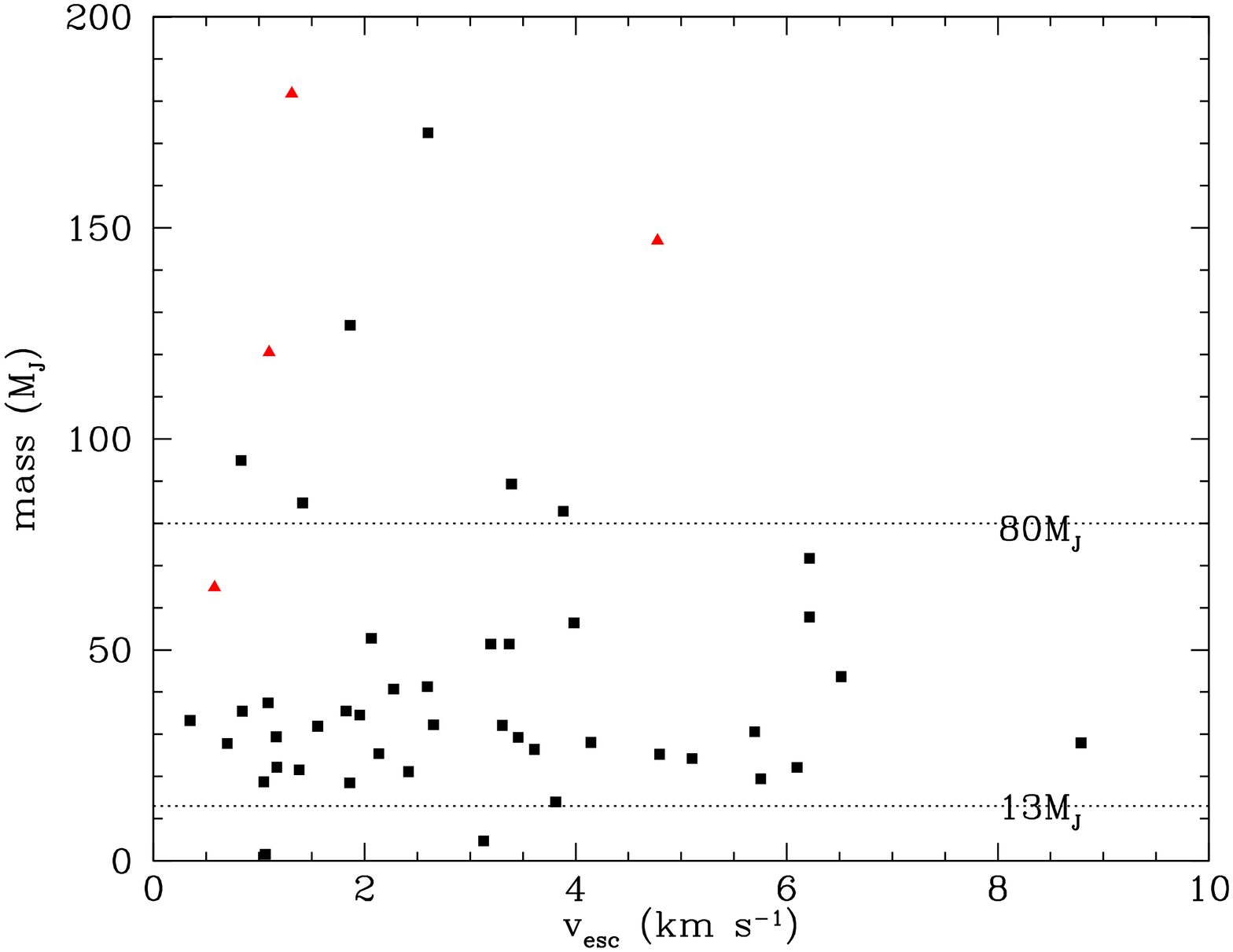}}
\caption{Top: the velocity distribution of the ejected stars; most stars are ejected with low velocities ($\la 3\,{\rm km}\,{\rm s}^{-1}$). Bottom: the ejection velocities plotted against mass; HBs tend to be ejected with somewhat lower velocities.}
\label{fig:ejectedvel}
\end{figure}
%%%%%%%%%%%%

\subsection{The ejected population}

More than half ($55\%$) of all the stars are ejected into the field. 9 (out of 29, i.e. $\sim 30\%$) of the HBs are ejected, and 45 (out of 67, i.e. $\sim 67\%$) of the BDs. The mass distribution of the ejected stars is shown in Fig.~\ref{fig:mspec.ejected}. Most of the objects ejected are low-mass BDs; there are also some PMs. 4 binaries are ejected, of which one comprises two HBs, one comprises an HB and a BD, one comprises two BDs, and one comprises a BD and a PM.

The velocity distribution of the ejected stars is shown in Fig.~\ref{fig:ejectedvel} (top). Most stars are ejected with low velocities ($\la 3\,{\rm km}\,{\rm s}^{-1}$). HBs tend to be ejected with somewhat lower velocities, but otherwise there is no strong dependence of the ejection velocity on the mass of the star (Fig.~\ref{fig:ejectedvel}, bottom). A small percentage of the ejected stars in our simulations have very high velocities (cf. Umbreit et al. 2005), and the lack of such high velocities in observational surveys of BD kinematics, and the apparent absence of a diaspora of brown dwarfs around young star clusters, has often been used as an argument against the ejection mechanism. However, in our simulations both BDs and HBs are ejected, and hence they should be roughly co-extensive (cf. Luhman 2004; Goodwin et al. 2005; Joergens 2006b). Moreover the ejection velocities would be significantly lower if the ejection were treated fully with hydrodynamics, rather than invoking sinks and then switching to an N-body code.

%%%%%%%%%%%%%%%
\begin{figure*}
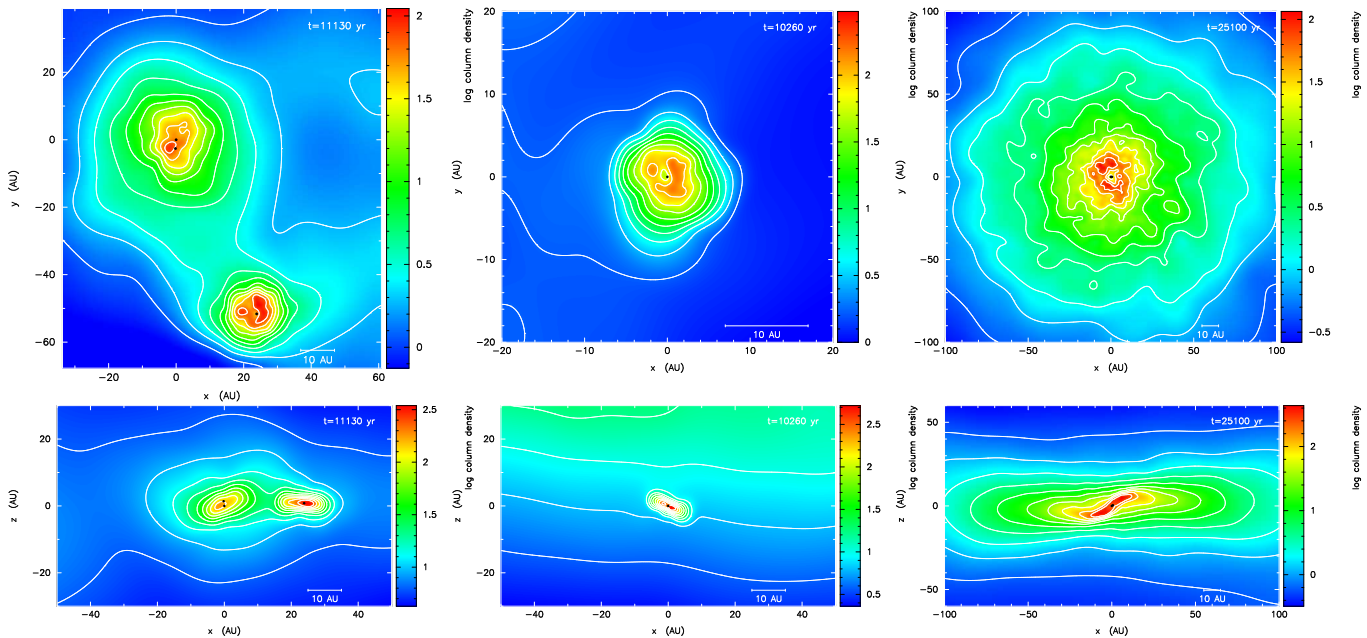

\centerline{
\includegraphics[height=6.2cm,angle=-90]{rdm02.sink6.xy.ps}\hspace{-0.4cm}
\includegraphics[height=6.2cm,angle=-90]{rdm11.sink4.xy.ps}\hspace{-0.4cm}
\includegraphics[height=6.2cm,angle=-90]{rdm08.sink2.xy.ps}}
\centerline{
\includegraphics[height=6.2cm,angle=-90]{rdm02.sink6.xz.ps}\hspace{-0.4cm}
\includegraphics[height=6.2cm,angle=-90]{rdm11.sink4.xz.ps}\hspace{-0.4cm}
\includegraphics[height=6.2cm,angle=-90]{rdm08.sink2.xz.ps}}
\caption{Rare systems produced by disc fragmentation. From left to right: (i) discs in a binary system which are neither aligned with each other nor with the ur-disc (the ur-disc is aligned with the $z=0$ plane), (ii) disc that is not aligned with the ur-disc, and (iii) a warped disc.}
\label{fig:nonaligned}\label{fig:warped}
\end{figure*}
%%%%%%%%%%%%%

\section{Rare systems formed by disc fragmentation}

There are four types of system formed in our simulations that are rare but nevertheless have similarities with intriguing observed systems.

\subsection{Free floating planetary-mass objects}

Free floating planetary-mass stars -- i.e. single field stars with masses $<13 {\rm M}_{_{\rm J}}$ -- have been observed in Orion (Lucas \& Roche 2000), in the $\sigma$ Orionis cluster (Zapatero Osorio et al. 2000) and in IC 348 (Najita et al. 2000). The simulations presented here show that such low-mass stars can form by disc fragmentation, and this is confirmed analytically by Whitworth \& Stamatellos (2006). It is uncertain whether they can also form from the collapse of low-mass cores (e.g. Greaves et al. 2005).
 
3 PMs form in our simulations, and all three of them are ejected into the field. Their masses remain below the D-burning limit because they are ejected from the disc very soon after formation. According to our simulations BDs should outnumber PMs by a factor of $\sim 14$. Allowing for the fact that 35\% of the BDs remain bound to the ur-star but all PMs are ejected, BDs should outnumber PMs in the field by a factor of $\sim 9$. However, we re-iterate that these numbers are for a very specific ur-system.

\subsection{Brown-dwarf/planetary-mass binaries}

One of the 13 low-mass binaries formed in our simulations has a $53\,{\rm M}_{\rm J}$ BD primary and an $11\,{\rm M}_{\rm J}$ PM secondary. This system has a semi-major axis of $110\,{\rm AU}$ and an eccentricity of 0.6. The two components of the system formed independently by disc fragmentation, paired up in the disc, and then were ejected as a binary into the field.  This system is qualitatively similar to 2MASS 1207-3932 (Chauvin et al. 2004, 2005), a system with a $25\,{\rm M}_{\rm J}$ BD primary orbited by a $5\,{\rm M}_{\rm J}$ PM secondary, with a  projected separation of $55\,{\rm AU}$. PMs are unlikely to be able to form by core accretion in BD discs (Lodato et al. 2005; Payne \& Lodato 2007).   Lodato et al. (2005) suggest that  2MASS 1207-3932B may have formed by fragmentation of a disc around 2MASS 1207-3932A. We suggest that such systems form by pairing up of two components that form indepedently in a fragmenting disc, capture one another, and then are ejected into the field as a binary.

\subsection{Systems with non-coplanar discs}

Four of the 58 circumstellar or circumbinary discs formed in our simulations are poorly aligned with the ur-disc (Fig.~\ref{fig:nonaligned}). This is because (a) stars do not form exactly on the midplane of the ur-disc, and (b) subsequently they experience impulsive perturbations due to passing stars that have formed in the disc. We presume that poorly aligned discs are even more common in real systems, i.e. asymmetric ur-discs forming in collapsing, fragmenting cores with lumpy material continuing to infall onto the ur-disc. The existence of poorly aligned discs is inferred from polarimetric observations (Jensen et al. 2004; Monin et al. 2006) and from non-parallel jets (e.g. Davis et al. 1994).

\subsection{Warped discs}

One of the discs formed in our simulations is noticeably warped (see Fig.~\ref{fig:warped}). The formation of such discs is a rare event, requiring a violent interaction between a star which is about to be ejected and a dense spiral arm. Warped discs have been observed (e.g. Heap et al. 2000; Quillen 2006).

\section{Implications for planet formation}
\label{sec:planets}

Our simulations show that disc fragmentation can produce PMs, but the probability is low; for every PM formed, there are more than 30 BDs and/or HBs formed. The PMs form at large distances from the ur-star, and are then ejected before they have time to grow by accretion. If they were not ejected, they would grow in mass by accretion, and end up above the D-burning limit. Thus, it is very hard, if not impossible, for planetary-mass stars to form in the outer reaches of massive extended ur-discs and then migrate inwards to a tight orbit around the ur-star.

Fragmentation is also unlikely to happen close to the ur-star (e.g. Matzner \& Levin 2005; Rafikov 2005; Whitworth \& Stamatellos 2006; Boley et al. 2006; Durisen et al. 2007; Stamatellos et al. 2007b; Stamatellos \& Whitworth 2008;  Cai et al. 2008).

Therefore exoplanets (e.g. Udry \& Santos 2007; {\it Extrasolar planets Encyclopaedia} at {\tt http://exoplanet.eu}) probably have to form by the core accretion mechanism (e.g. Safronov  1969; Goldreich \& Ward 1973; Pollack et al. 1996), at later stages during disc evolution. In fact even this may be rather hard in the ur-discs we have simulated here, although not necessarily in other discs. This is because each ur-disc we have simulated ends up with a low-mass star (usually an HB) on a close ($\la 30\,{\rm AU}$) eccentric orbit, and this is likely to inhibit planet formation in the inner disc (Th{\`e}bault et al. 2006)

\section{Conclusions}

The fragmentation of massive, extended ur-discs  produces stars which populate the low-mass end of the stellar initial mass function: low-mass H-burning stars (HBs), brown-dwarf stars (BDs) and planetary-mass stars (PMs). Despite the fact that we have considered only a single ur-system, the predictions of the model compare well with observation. Simulations of different ur-systems (different ur-disc masses, radii, surface-density profiles, etc. and different ur-star masses) yield similar results (Stamatellos et al., in prep), as do simulations in which we follow the formation of the ur-system from a collapsing turbulent prestellar core (Attwood et al., in prep).

For the particular ur-system that we simulate (a $0.7\,{\rm M}_\odot$ $400\,{\rm AU}$ disc in orbit about an $0.7\,{\rm M}_\odot$ star), we predict the following.

\begin{itemize}

\item{Each ur-disc fragments to form between 5 and 11 stars; 30\% are HBs, 67\% are BDs and 3\% are PMs.}

\item{55\% of the stars are ejected into the field (30\% of the HBs, 65\% of the BDs and 100\% of the PMs).}

\item{The ur-star ends up with a close companion (usually an HB) and two or three wide companions (typically an HB and one or two BDs).}

\item{These wide companions may include a low-mass binary system.}

\item{Within a few thousand years the ur-disc is reduced to $M_{_{\rm DISC}}\la 0.01\,{\rm M}_\odot$ and $R_{_{\rm DISC}}\la 100\,{\rm AU}$.}

\item{We hypothesise that there is a range of masses (say between $\sim 0.05\;{\rm and}\;\sim 0.2\,{\rm M}_\odot$) where -- {\it proceeding to lower masses} -- the balance gradually shifts {\it from} the majority of stars forming by turbulent fragmentation (as the ur-star presumably did), {\it to} the majority of stars forming as the result of fragmentation of discs around such stars (like the ur-disc).}

\item{Only a small fraction (say $20\;{\rm to}\;30\,\%$) of the higher-mass Sun-like stars (G, K and early M dwarfs) need to support disc fragmentation of the type we have modelled here to supply most of the low-mass HBs, BDs and PMs that are observed.}

\item{Most stars initially condense out of the ur-disc between $\sim 100\,{\rm AU}$ and $\sim 200\,{\rm AU}$.}

\item{The stars forming closer to the ur-star tend to grow to become HBs, and may migrate inwards to much closer final orbits.}

\item{The stars forming further out tend to migrate outwards and/or be ejected.}

\item{As a result there is a brown dwarf desert, i.e. a lack of brown dwarfs orbiting close to the ur-star; those brown dwarfs which are not ejected tend to end up orbiting out beyond $\sim 200\,{\rm AU}$.}

\item{Due to dynamical interactions between stars, stellar orbits can end up quite eccentric and strongly inclined to the plane of the ur-disc.}

\item{The majority of stars formed from the ur-disc have their own accretion discs, and some of these discs are retained during ejection; however, we expect BDs that remain bound to the ur-star to have a higher disc frequency than those which are ejected.}

\item{Stars formed in the ur-disc frequently pair up to form low-mass binary systems. The low-mass binary frequency is $\sim 16\%$.}

\item{These binary systems are usually close, but can be wide; they tend to have eccentric orbits and large mass ratios (i.e. components of comparable mass); they include HB/HB pairs, HB/BD pairs, BD/BD pairs and BD/PM pairs.}

\item{These binary systems can survive ejection; however, a BD that remains bound to the ur-star is three to five times more likely to be in a BD/BD binary system than a BD in the field.}

\item{Ejection velocities are $\ga 3\,{\rm km}\,{\rm s}^{-1}$, but these would probably be reduced if the hydrodynamics of ejection were modelled properly (i.e. not invoking sinks and not switching to an N-body code).}

\item{Hot Jupiters are very unlikely to form by disc fragmentation. Moreover in discs which fragment like the ones modelled here, formation of Hot Jupiters by core accretion may also be inhibited by the stellar companions that form on close eccentric orbits.}

\end{itemize}

We conclude that disc fragmentation is a robust mechanism for the formation of brown dwarf stars, as well as planetary-mass stars and low-mass H-burning stars. It explains successfully properties that are not satisfactorily explained by other formation mechanisms, for example the brown dwarf desert and the observed statistics of low mass binary systems. We suggest that a large proportion of brown dwarf stars and planetary-mass stars may be formed by disc fragmentation

\section*{Acknowledgements}
%%%%%%%%%%%%%%%%%%%%%%%%%%%
   
We would like to thank the referee for his helpful comments. The computations reported here were performed using the UK Astrophysical Fluids Facility (UKAFF). N-body simulations were done by a code originally developed by D. Hubber, who we thank for the support provided. We also thank Ken Rice, Aaron Boley,  \& Viki Joergens for useful discussions/suggestions. We also gratefully acknowledge the support of an STFC rolling grant (PP/E000967/1) and a Marie Curie Research Training Network (MRTN-CT2006-035890).

\label{lastpage}

\end{document}